\documentstyle[psfig,times,mathptm,epsfig,here,12pt,twoside]{mbook}
\begin{document}

\pagestyle{empty}

\begin{center}

{\Huge\    JAGELLONIAN UNIVERSITY         \\
            INSTITUTE OF PHYSICS          \\
}

\vspace{1.75cm}

\begin{figure}[H]
\centerline{\parbox{0.3\textwidth}{\epsfig{file=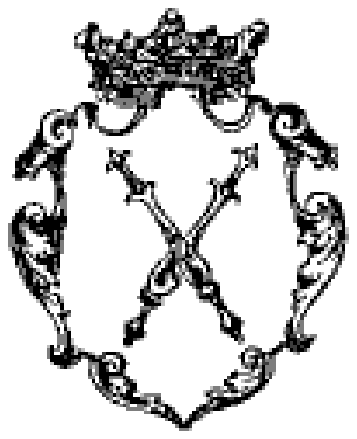,width=0.35\textwidth,angle=0.}}}
\end{figure}

\vspace{1.75cm}

{\bf{\LARGE\   Bremsstrahlung radiation in the \\
               deuteron - proton collision
}}

\vspace{0.5cm}

{\large                 Joanna Przerwa}

\vspace{2.0cm}

\end{center}

\vspace{1.5cm}

\begin{flushright}
Master Thesis
prepared at the Nuclear Physics Department\\ 
guided by: dr. Pawe{\l} Moskal  \\
\end{flushright}
\vspace{0.5cm}
\begin{center}
    Cracow 2004
\end{center}


\newpage
\pagestyle{empty}
\begin{center}

\end{center}

\vspace{15.cm}

\begin{flushright}
\textbf{ PER ASPERA AD ASTRA}
\end{flushright}


\newpage
\pagestyle{empty}
\begin{center}

\end{center}

\begin{flushright}
\large{\it
It is my pleasure to express my gratitude to a large number\\
of people without whom this work wouldn't have been possible.\\

\vspace{0.5cm}

First of all I thank dr. Pawe{\l} Moskal --- a person
who greatly influenced my life and career --- for encouraging
me to succeed in achieving high goals. He was always and is
now an example to follow as a scientist and educator.

\vspace{0.5cm}

I would like to express my profound gratitude to Prof. Walter Oelert
for giving me the opportunity to work within the COSY-11 group. I am also indebt
for comments, valuable advice and encouragement.\\

\vspace{0.5cm}

Less direct but not less important has been the inspiration of\\
Prof. Lucjan Jarczyk and Prof. Bogus{\l}aw Kamys.\\

\vspace{0.5cm}

I am also very grateful to Prof. Reinhard Kulessa for allowing me to prepare 
this thesis in the Nuclear Physics Department of the Jagellonian University.\\

\vspace{0.5cm}

I want to express my appreciation to all Colleagues from the COSY-11 group
with whom I have the great fortune to interact and work.\\

\vspace{0.5cm}

I also thank my colleagues: Micha{\l} Janusz, Ma{\l}gorzata Kasprzak,
Marcin Ku{\'z}niak, and Tytus Smoli{\'n}ski for the nice atmosphere during
work.\\

\vspace{0.5cm}

The last, but not least, is my gratitude to my mother, who gave me\\ 
the
courage to get my education and supported me in all achievements.\\
Many thanks to my brother Micha{\l} --- you are the second half of my brain !
} 
\end{flushright}


\newpage
\pagestyle{empty}

\begin{center}
{\bf Abstract\\

\vspace{1.0cm}

Bremsstrahlung radiation in the deuteron -- proton collision}
\end{center}

\vspace{1.0cm}

   Despite the fact that Bremsstrahlung radiation has been observed many
   years ago, it is still the subject of interest of many theoretical and
   experimental groups. Due to the high sensitivity of the $NN~\to~NN\gamma$
   reaction to the nucleon~--~nucleon potential, Bremsstrahlung radiation is used
   as a tool to investigate details of the nucleon~--~nucleon interaction.
   Such investigations
   can be performed at the cooler synchrotron COSY in the Research Centre
   J{\"u}lich, by dint of the COSY--11 detection system.\\
   For the first time at the COSY--11 experiment  signals from  $\gamma$ -- quanta
   were observed in the time -- of --  flight distribution of neutral particles measured
   with the neutral particle detector.\\
   In this thesis the results of the identification of Bremsstrahlung radiation
   emitted via the $dp~\to~dp\gamma$ reaction
   in data taken with a proton target and a deuteron beam
   are presented and discussed.\\
   The time resolution of the neutral particle detector and its timing calibration
   are crucial for the identification of the $dp~\to~dp\gamma$ reaction.
   Therefore, methods of determining the relative timing between individual
   modules -- constituting
   the neutron detector -- and of the general time offset
   with respect to the other detector components are described.
   Furthermore the accuracy of the momentum determination of the registered
   neutron which defines the precision of the event reconstruction was extracted
   from the data.

\tableofcontents
\newpage
\chapter{Introduction}

\pagestyle{myheadings}
  \markboth{Introduction}{Introduction}
In collisions between nucleons electromagnetic radiation can
be emitted due to the rapid change of the nucleon velocity.
This radiation is referred to as  bremsstrahlung radiation.
Although it has been observed many years ago,
it is still the subject of interest of many experimental and theoretical groups~\cite{khokholov,cozma,bilger,greiff}.
Due to the high sensitivity of the $NN~\to~NN\gamma$ reaction
to the nucleon--nucleon potential, bremsstrahlung radiation is used as a tool
to investigate details of the NN interaction and to discriminate between various
potential models~\cite{khokholov,cozma}.
\par
Such investigations can be performed at the cooler synchrotron COSY in the Research~Center
J{\"u}lich, by means of the facilities installed there
like the internal COSY--11 detection setup. Although the COSY--11 facility is designed for
threshold meson production covering a small solid angle in the laboratory system
the special configuration allows also the study of reaction channels at high excess energies.
The main aim of this thesis is an identification of bremsstrahlung radiation
in a data sample taken by the COSY--11 collaboration during a measurement in January 2003~\cite{smyrski,Cezary}.
The measurement was carried out at a hydrogen
cluster target~\cite{dombrowski} using a deuteron beam with a momentum of 3.204~GeV/c.
The registration of gamma quanta
became possible because the COSY--11 detection system has been
extended by a neutral--particle--detector.
This detector  enables not only to measure the bremsstrahlung radiation created in
the collision of nucleons but also opens wide possibilities to investigate the isospin
dependence of the meson production in the hadronic interactions~\cite{moskal-review}.
For example at present the COSY--11 facility permits
to investigate $\eta$--meson production in proton--proton and proton--neutron collisions~\cite{moskal-hadron,janusz}.
Neutrons and gamma quanta registered in the neutral--particle--detector are identified via the time--of--flight
on the 7 m
distance between the target and the detector.
In case of a neutron the time--of--flight and the hit position enables to
determine its four--momentum vector. Since the time resolution of the detector and its timing calibration are
crucial for the identification of the studied reactions, the data analysis of the $dp~\to~dp\gamma$ reaction
will be preceded by the detailed presentation of the method used for the time calibration of this detector.

\par
The present thesis is divided into six chapters. The second chapter --- following the introduction
--- describes briefly the motivations to investigate the $NN~\to~NN\gamma$ process, in particular,
in view of the study of the proton--$\eta$ interaction by the COSY--11 group.\\
\par
Description of the detection system, with a special emphasis on the structure
of the neutron detector and the basis of its functioning will be presented in
the third chapter.\\
\par
The fourth chapter is devoted to the time calibration of the neutral--particle--counter.
The method of
determining the  relative timing between the individual modules and the general time offset with respect
to the other detectors will be depicted. This chapter describes also the determination of the
momentum resolution of the
registered neutrons.\\
\par
The results of the identification of the $dp~\to~dp\gamma$ reaction
in data taken in January 2003
are presented in the fifth chapter, where also a detailed description of the data analysis is included. \\
\par
The sixth chapter  comprises summary and perspectives, in particular a possibility to study the pentaquark
state $\Theta^{+}$ at the COSY--11 facility is discussed.

\chapter{Motivation}
\pagestyle{myheadings}

\markboth{\bf Motivation}
         {\bf Motivation}

For many years experimental and theoretical studies have been devoted to
distinguish among various potential
models and estimate of the  off-shell amplitudes.
The idea of using nucleon--nucleon bremsstrahlung as a tool for investigating
this problem has attracted attention since calculations of the cross sections of the $NN~\to~NN\gamma$
reaction are highly sensitive to the
$NN$ potential.
\par
We would like to study this process also in view of another important issue related to the
$\eta$--proton interaction.

\begin{figure}[H]
\centerline{\parbox{0.6\textwidth}{\epsfig{file=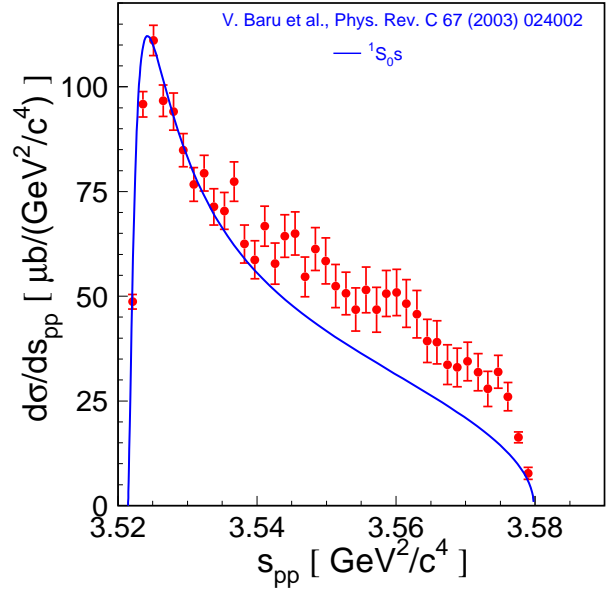,width=0.65\textwidth,angle=0.}}}
       {\caption{
                    Differential cross section for the $p~p~\to~p~p~\eta$
                    reaction as a function of the invariant mass of
                    proton--proton system. Data are from ref.~\cite{moskal-ppdynamics}
                    and the line corresponds to the model of reference~\cite{baru}.
                    Picture adapted from~\cite{moskal-ppdynamics}.
 }}
\end{figure}
Figure 2.1 shows a differential cross section for the $pp~\to~pp\eta$
reaction as a function of the invariant mass of the proton--proton system.
The measurement  has been performed at an excess energy
of Q~=~15.5~MeV. The theoretical description
does not agree with data in the whole
range of invariant mass of the proton--proton system. It has been pointed out
that the difference between the  data and the theoretical predictions
originates from  proton--$\eta$ interaction in the final state.
In order to prove that this conclusion does not depend on the applied model it would be
desirable to compare this distribution with the one obtained for the $ppX$ system,
where $X$ does not interact strongly with protons. Therefore best suited for this
purpose would be the $pp\gamma$ system.
Let us assume, that the differential cross section for the $pp~\to~pp\gamma$
reaction as a function of proton--gamma invariant mass are determined.
If the theoretical model, used for the calculations presented by the solid line in figure
2.1 were in perfect agreement with those data, it would corraborate the assertion that
the disagreement between data and theory in the $pp\eta$ case
is due to the p--$\eta$ interaction.\\
Studies of the $pp$--bremsstrahlung by means of the COSY--11 facility haven't been performed,
but production of bremsstrahlung radiation  in proton--proton collisions
was investigated at the COSY--TOF facility~\cite{bilger}.
Data have been taken at a proton beam momentum of 797~MeV/c using a wide angle spectrometer.
At the COSY--11 facility as a first step of investigations of the bremsstrahlung radiation the
$dp~\to~dp\gamma$
reaction is studied, and
the main point of the present work is an identification of dp$\gamma$ events
in data taken during the experiment devoted to the measurement of the $\eta$--meson production
in deuteron--proton collisions.

\vspace{-1.0cm}

\begin{figure}[H]
\centerline{\parbox{0.7\textwidth}{\epsfig{file=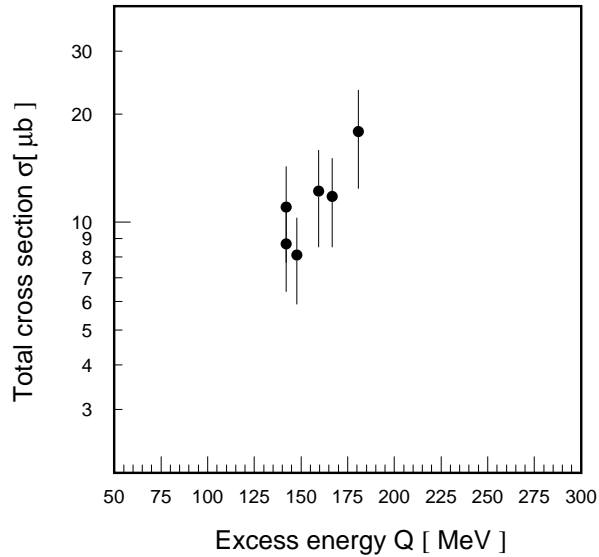,width=0.75\textwidth,angle=0.}}}
\vspace{-1.0cm}
       {\caption{
                    Total cross section for the $dp~\to~dp\gamma$
                    reaction as a function of the excess energy.
                    Original data are taken from~\cite{greiff}.
 }}
\end{figure}
A similar experiment --- measurement of the  $dp~\to~dp\gamma$
reaction --- has been performed by the WASA/PROMICE collaboration at the storage ring CELSIUS,
with deuteron beam energies between 437~MeV and 559~MeV. In these studies angular and spectral distributions
are divided into two groups that can be attributed to a quasifree $np~\to~np\gamma$ process,
and to a gamma production process with all three nucleons involved, viz: $dp~\to~dp\gamma$.
The results of these investigations ---
the total cross section for the $dp~\to~dp\gamma$ reaction --- are given in  figure 2.2.
\par
In this work the feasibility of the bremsstrahlung measurement at the COSY--11
detection setup will be proven,
and as soon as the analysis is finished the total cross section data--base for the $dp~\to~dp\gamma$ reaction
(fig. 2.2) will be extended by one point at Q~=~557~MeV.

\chapter{ Experimental setup}
\markboth{\bf Experimental setup}
         {\bf Experimental setup}

\section{COSY--11 facility---general remarks}

The experiment described in this thesis has been performed at the COSY--11 facility~\cite{brauksiepe},
an internal magnetic spectrometer installed at the cooler synchrotron COSY in J{\"u}lich~\cite{maier}.
The COSY--11 detection system is schematically depicted in figure 3.1. Details of the functioning
of all detector components and the method of measurement can be found in references~\cite{brauksiepe,moskal-dt,wolke-dt}.
Therefore, here the used experimental technique will be only briefly presented.
\par
The synchrotron accelerates protons
and deuterons up to a momentum of 3.4~GeV/c. At the highest momentum a few $10^{10}$ accelerated particles pass
through the target $\sim 10^6$ times per second.\\
The hydrogen ($H_{2}$) or deuteron ($D_{2}$) cluster target (see figure 3.2)
is installed in front of the dipole magnet. The positively charged products of the reaction are bent in the
magnetic field of the dipole and leave the vacuum chamber through thin exit foils,
whereas the beam --- due to the much larger momentum --- remains on its orbit inside the ring.
The charged ejectiles are detected in drift chambers (D1, D2, D3)~\cite{brauksiepe}
and scintillator hodoscopes (S1, S2, S3)~\cite{brauksiepe,moskal-dt}.
Neutrons and gamma quanta are registered in the neutron detector (N). In order to separate
neutrons and gamma quanta from charged particles veto detector (V) is used.
An array of silicon pad detectors ($Si_{spec}$)
is used for the registration of the spectator protons.
Protons scattered under large angle are measured in another position sensitive silicon
pad detector $(Si_{mon})$.
\par
The experiments performed at COSY--11 base on the measurement of four-momenta of the outgoing particles.
Unregistered short lived mesons and hyperons are identified via the missing mass technique.
\par
For each charged particle, which gave  signals in  drift chambers, the momentum vector
can be determined. First the trajectories of
the particles are reconstructed~\cite{sokolowski}, and then knowing the magnetic field of the dipole,
the momentum vector is reconstructed.
In case of two close tracks, the information about the energy loss from S1, S2, and S3 is used to inspect
the efficiency of the track reconstruction.
Particle's velocity determination is based on the time-of-flight measurement between S1 (or S2)
and S3 detectors. Knowing the velocity and the momentum of the particle, its mass can be calculated,
and hence the particle can be identified. After the particle identification the time of the reaction
at the target is obtained from the known trajectory, velocity, and the time measured
by the S1 detector.
The neutron detector delivers the information about the time at which the registered neutron or gamma
quanta induced a hadronic or electromagnetic reaction.
\newpage
\begin{center}
\end{center}

\vspace{1.0cm}

\begin{figure}[H]
\centerline{\parbox{0.9\textwidth}{\epsfig{file=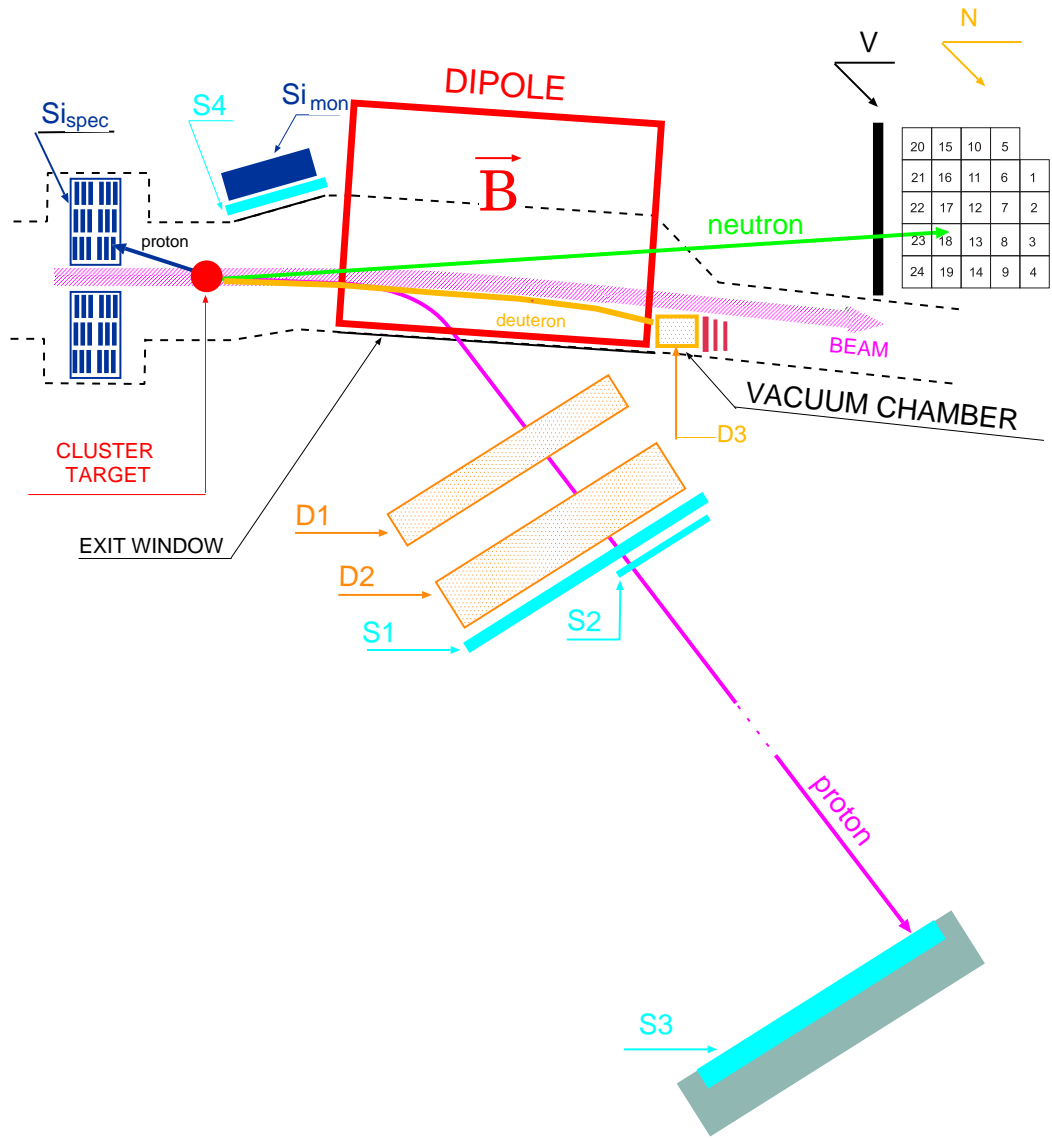,width=0.95\textwidth,angle=0.}}}

\vspace{1.0cm}
        {\caption{
                     Scheme of the COSY--11 detection system.
                     Protons are registered in two drift chambers D1, D2
                     and in the scintillator hodoscopes S1, S2, S3.
                     An array of silicon pad detectors $( Si_{spec})$
                     is used for the registration of the spectator protons.
                     Neutrons are registered in the neutron modular detector (N).
                     In order to distinguish neutrons from charged particles a veto
                     detector (V) is used. Deuterons with larger momentum are registered in deuteron chamber D3.
 }}
\end{figure}

\begin{figure}[H]
\centerline{\parbox{0.45\textwidth}{\epsfig{file=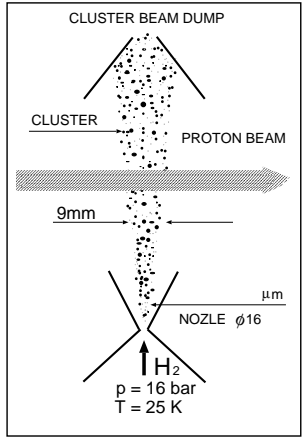,width=0.6\textwidth,angle=0.}}}
        {\caption{  Schematic view of the cluster target.
 }}
\end{figure}

The time of the reaction combined with this
information allows to calculate the time--of--flight (TOF$^N$) of the neutron (or gamma) between the target
and the neutron detector, and --- in case of neutrons --- to determine the absolute
value of the momentum (p) what can be expressed as:
\begin{equation}
 p = m \cdot {l \over {TOF^N}} \cdot  {1 \over {\sqrt{1 - ({l \over {TOF^N}})^2}}},
\end{equation}
where m denotes the mass of the particle, $l$ stands for the the distance between the target and the
neutron detector and $TOF^N$ is the time--of--flight of the particle.
\par
In order to calculate the four--momentum of the spectator proton --- in case of quasi--free
reactions with proton beam and deuteron target --- its kinetic energy (T) is directly
measured as the energy loss in the silicon detector $( Si_{spec})$. Knowing the proton kinetic
energy one can calculate its momentum using the following relationship:
\begin{equation}
     p = \sqrt{{(T + m)}^2 - m^2},
\end{equation}
where $m$ denotes the proton mass, and $T$ is the kinetic energy.
In case of the measurements with deuteron beam and proton target the
trajectory of the spectator proton may be reconstructed from signals
from the drift chambers D1 and D2 and hence in this case its momentum can also
be determined.
\par
To evaluate the luminosity, the
elastically scattered protons are measured at the same time.
With one proton detected in the drift chambers and scintillator hodoscopes
and the other proton in the silicon detector $Si_{mon}$ resulting in the determination of the hit position,
the elastically scattered protons can be well separated.

\section{Functioning of the neutron detector}

In this section I would like to emphasize the neutron detector, since the time
calibration of this detector constitutes one of the main goals of this thesis.
\par
Previously, the neutron detector was built out of 12 detection units,
with light guides and photomultipliers mounted on one side of the module.
In order to improve the time resolution of the detector additional light guides
and photomulipliers were installed, such that the light signals from scintillation layers
are read out at both sides of the module. The present neutron detector consists of 24 modules,
like the shown in figure 3.3. Each module is built out of eleven plates of
scintillator material with dimensions 240 mm x 90 mm x 4 mm interlaced with
eleven plates of lead with the same dimensions. The scintillators are read out
at both edges of the module by light guides --- made of plexiglass --- whose
shape changes from rectangular to cylindrycal, in order to accumulate the produced light on the circular
photocathode of a photomultiplier.\\
\begin{figure}[H]
\centerline{\parbox{0.75\textwidth}{\epsfig{file=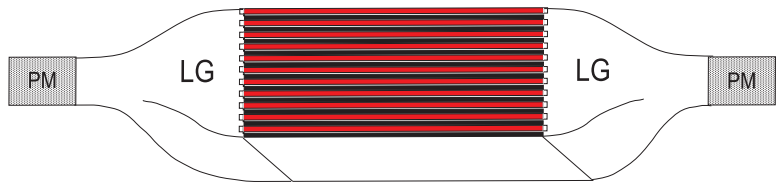,width=0.7\textwidth,angle=0.}}}
\vspace{-0.1cm}
         {\caption{  Schematic view of a single module of the neutron detector.
                    LG and PM denote light guides and photomultipliers, respectively.
                    Picture is taken from~\cite{czyzyk-mgr}.
 }}
\end{figure}
\vspace{-0.80cm}
\begin{figure}[H]
\centerline{\parbox{0.55\textwidth}{\epsfig{file=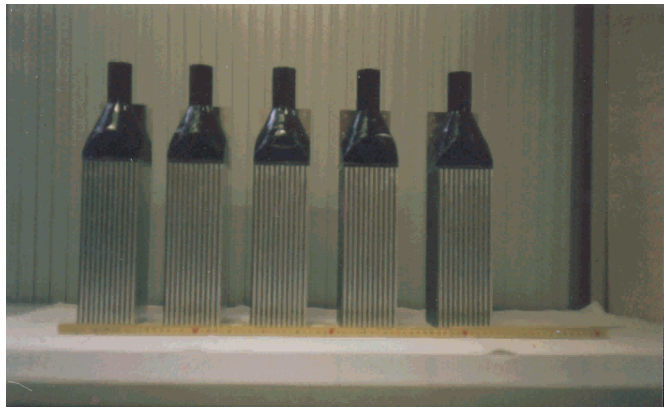,width=0.6\textwidth,angle=0.}}}
        {\caption{ Photo of single modules of the neutron detector, at the stage when the
                   scintillator/lead structure was well visible.
 }}
\end{figure}
The neutron detector is positioned at a distance
of 7 m behind the target in the configuration schematically depicted in figure 3.5.
 As can be deduced from figure 3.6
the maximum efficiency, for a given total thickness, for the registration of the neutron --- in the kinetic energy
range of interest for the COSY--11 experiments ($\sim$ 300~MeV -- $\sim$ 700~MeV) ---  would be achieved for the homogeneous
mixture of lead and scintillator. In order to optimize the efficiency and the cost
of the detector the plate thickness has been chosen to be 4 mm. This
results in an efficiency which is only few per cent smaller than the maximum possible.

\begin{figure}[H]
\centerline{\parbox{0.50\textwidth}{\epsfig{file=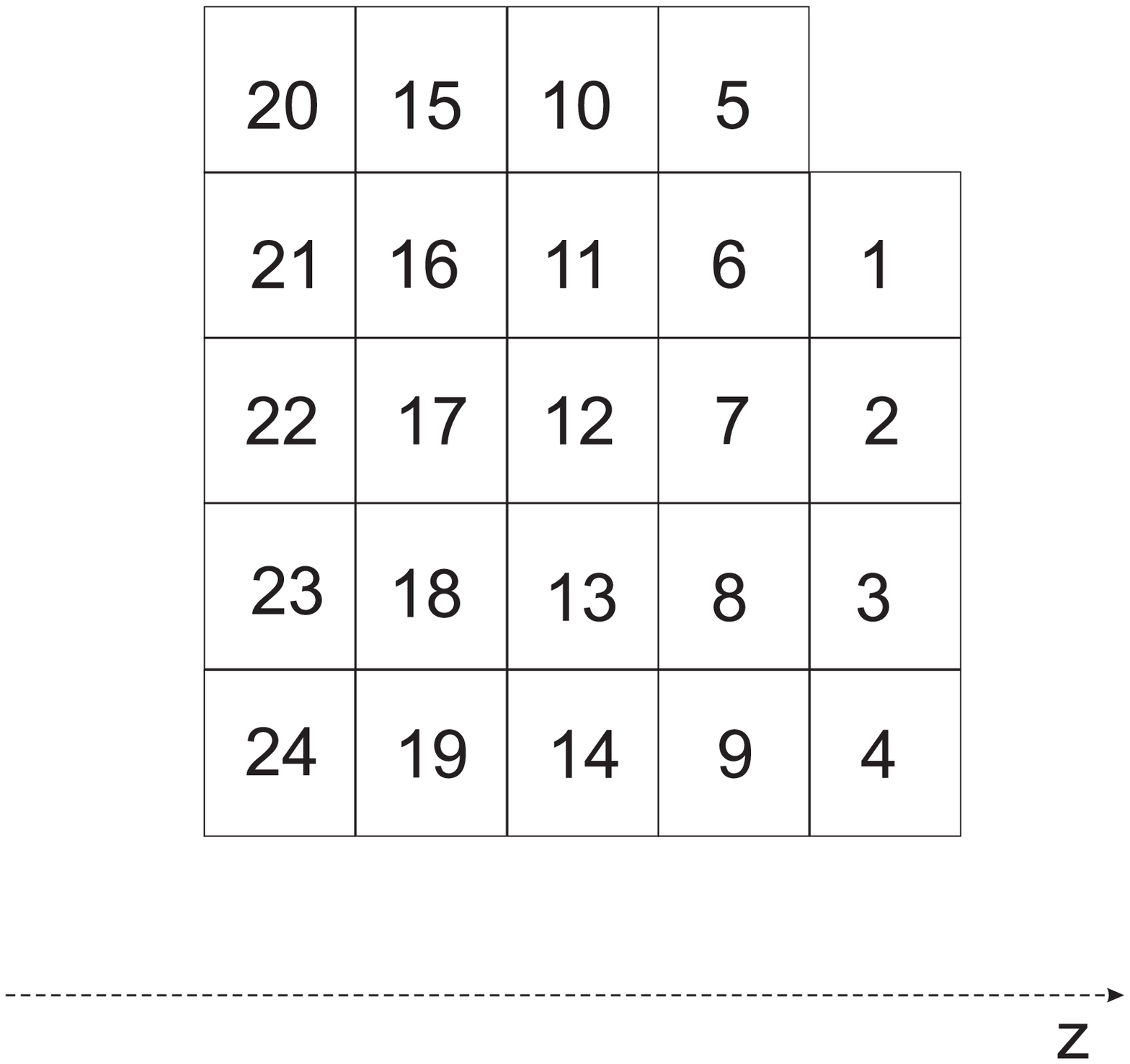,width=0.45\textwidth,angle=0.}}}
        {\caption{ Schematic view of the neutron detector from above.
                   The squares with numbers represent single detection modules.
                   The z axis is defined by the beam direction.
 }}
\end{figure}

\begin{figure}[H]
\centerline{\parbox{0.75\textwidth}{\epsfig{file=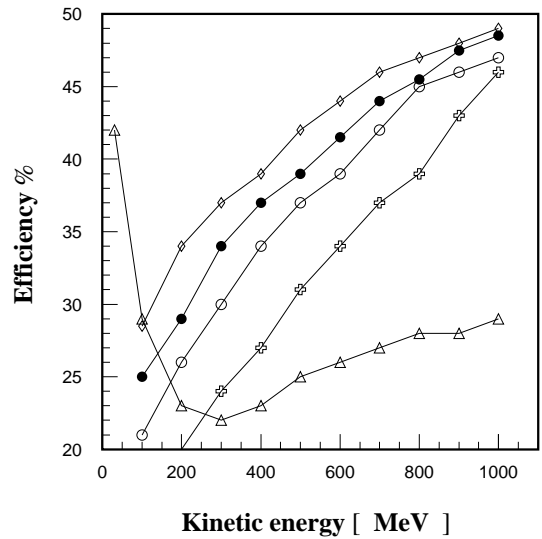,width=0.7\textwidth,angle=0.}}}
\vspace{-1.0cm}
        {\caption{ Calculated efficiencies of the 20 cm thick neutron detector as
                   a function of neutron kinetic energy for various layer
                   thicknesses of the scintillator and lead plates.
                   The solid line with triangles is for pure scintillator.
                   Diamonds denote a layer thickness of 0.5 mm, closed circles of 2 mm,
                   open circles of 5 mm, and crosses of 25 mm. The figure has been adapted
                   from~\cite{land}.
 }}
\end{figure}

The functioning of the detector was already confirmed in experiments
carried out with a deuteron beam and hydrogen target.
Figure 3.7 shows experimental distributions of the number of hits per individual
detection unit, for neutral (left)
and charged particles (right). As expected the counting rate of modules in the middle part of the
detector is much higher than these for the modules on the edges. In particular the smallest rate
is observed in modules No. 5, 10, 15, and 20, because this row is partly out of the
geometrical acceptance of the dipole yoke.
The number of hits per segment from experiment is in perfect agreement with
the corresponding spectra which were simulated  using a GEANT--3 code (see figure 3.8).

\vspace{-1.0cm}

\begin{figure}[H]
\parbox{0.45\textwidth}{\hspace{-0.5cm}\epsfig{file=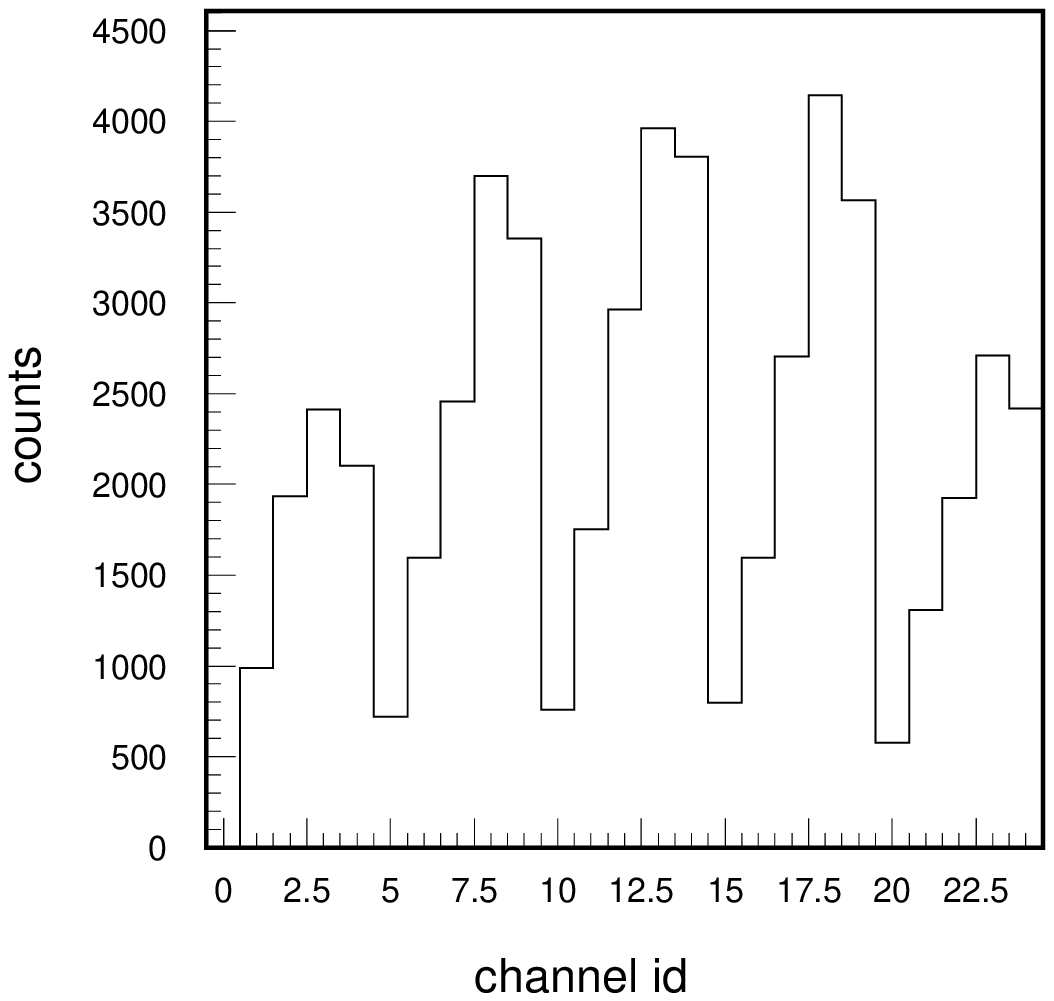,width=0.59\textwidth,angle=0.}}
\parbox{0.45\textwidth}{\hspace{-1.5cm}\epsfig{file=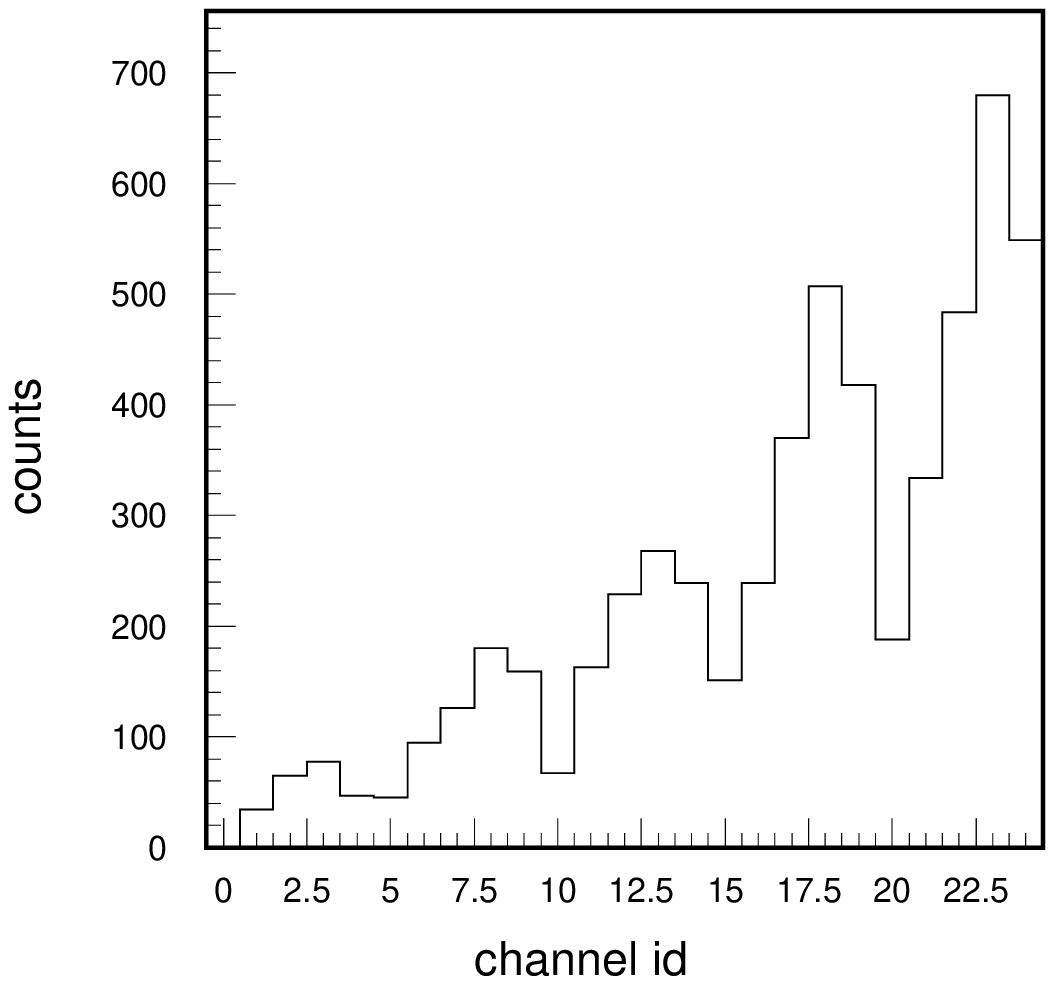,width=0.59\textwidth,angle=0.}}
\vspace{-1.5cm}
       {\caption{  Number of hits per individual module. Experimental spectra
                   obtained for neutral particles (left) and charged particles
                   (right) are shown.
 }}
\end{figure}

\vspace{-1.5cm}

\begin{figure}[H]
\parbox{0.45\textwidth}{\hspace{-0.5cm}\epsfig{file=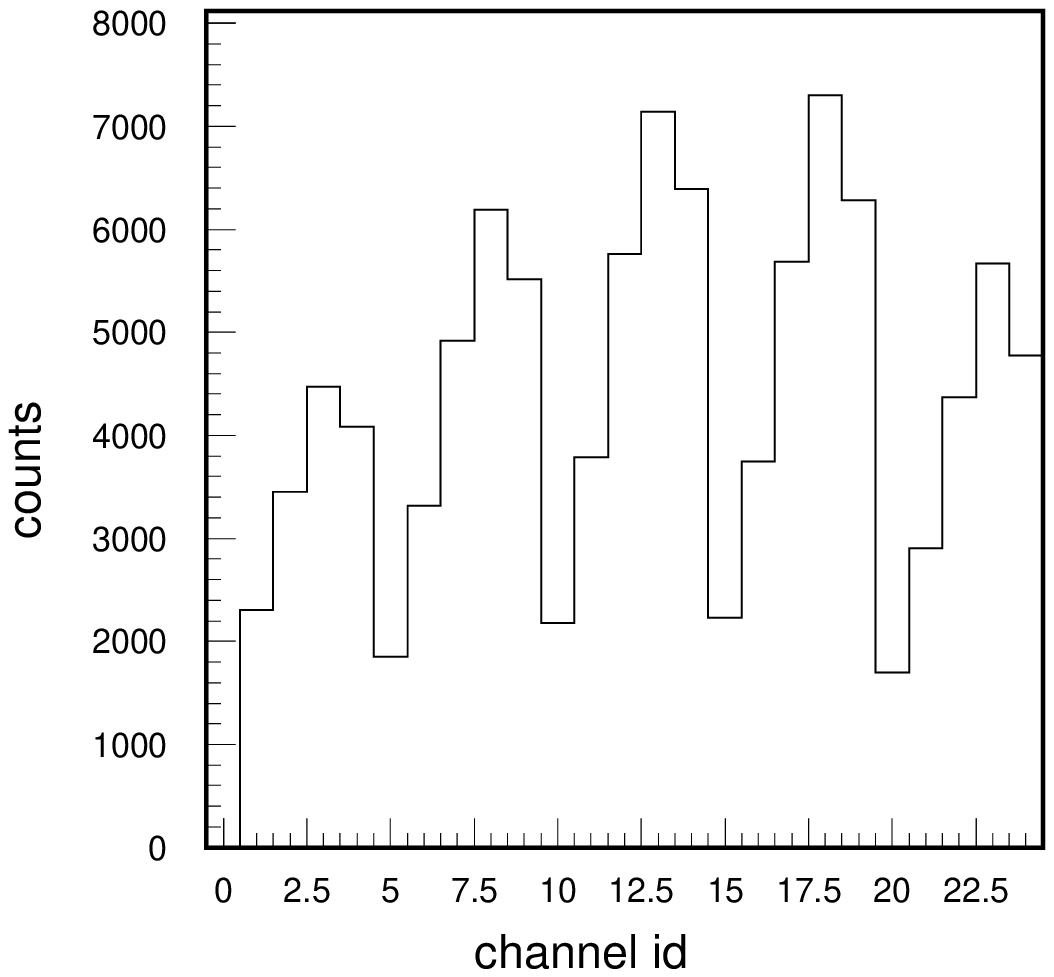,width=0.59\textwidth,angle=0.}}
\parbox{0.45\textwidth}{\hspace{-1.5cm}\epsfig{file=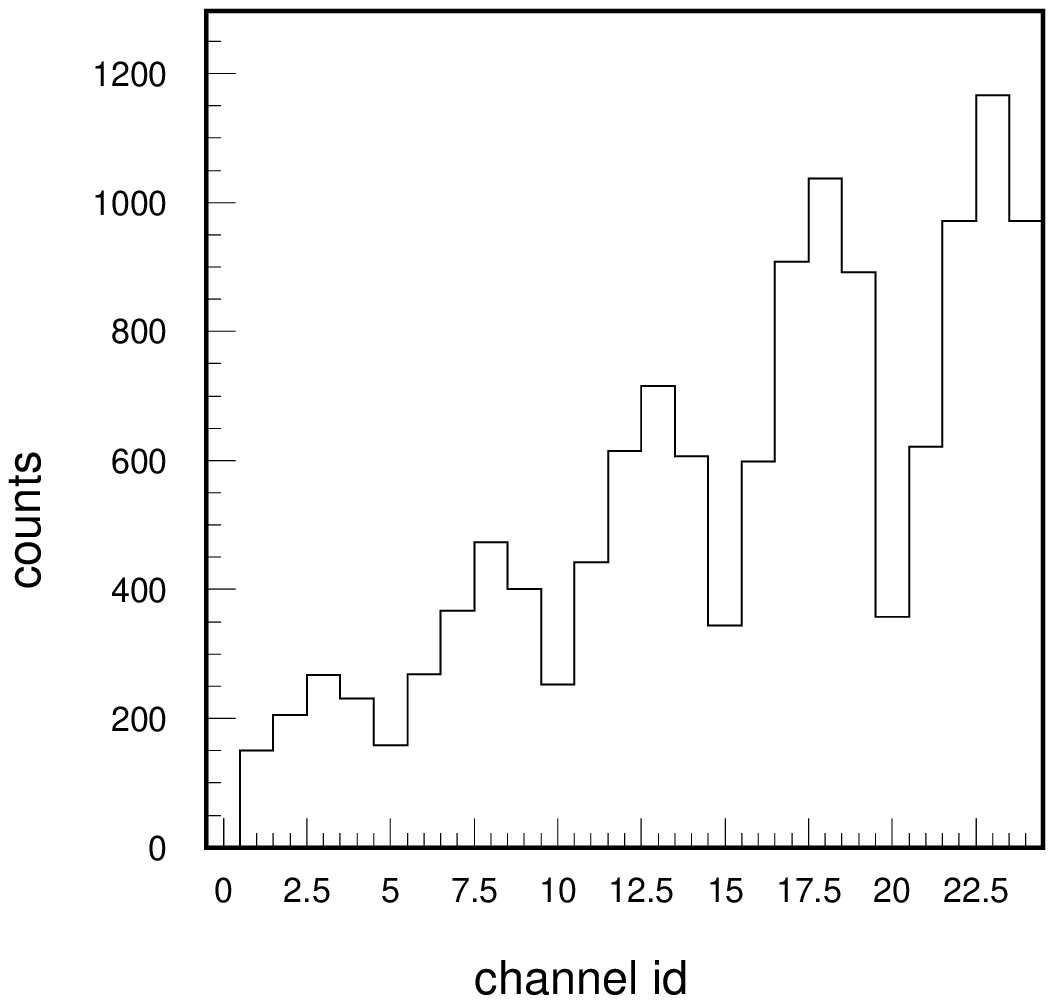,width=0.59\textwidth,angle=0.}}
 \vspace{-1.5cm}
       {\caption{  Number of hits per individual module from Monte-Carlo studies
                   obtained for neutral particles (left) and charged particles
                    (right). The simulation was performed for the $dp~\to~ppn_{sp}$
                    reaction taking into account reactions of the ejectiles with the dipole
                    and all detector materials.
 }}
\end{figure}

\chapter{Calibration of the neutron detector}
\markboth{\bf Calibration of the neutron detector}
         {\bf Calibration of the neutron detector}
The installation of the neutral particle detector at the COSY--11
facility enables to study a plethora of new reaction channels. This detector
is designed to deliver the time at which the registered neutron or gamma quantum
induced a hadronic or electromagnetic reaction, respectively. This information combined with
the time of the reaction at target place --- deduced using
other detectors --- enables to calculate the energy of the detected neutron.
In this section a method
of time calibration will be demonstrated and results achieved by its application
will be presented and
discussed. Information about the deposited energy is not used in the data analysis
because the smearing of the neutron energy determined in this manner is by more than
order of magnitude larger than this established from the time--of--flight method.
\vspace{-2.0cm}
\section{Time calibration of the neutron detector}
\subsection{Time signals from a single detection unit}
As already discussed in section 3.2 the neutron detector at the
COSY--11 facility is built out of 24 modules such as the one
shown in the figure below.

The time ($T^{exp}$) from a single module is calculated as an average time measured
by the upper and lower photomultiplier. Namely:
\begin{equation}
   T^{exp} = {{{T}^{up}_{TDC} + {T}^{dw}_{TDC}} \over 2 },
\end{equation}
where $T_{TDC}$ denote the time difference between the arrival of the
photomultiplier and trigger signals to the Time--to--Digital--Converter (TDC).

\begin{figure}[H]
\centerline{\parbox{0.75\textwidth}{\epsfig{file=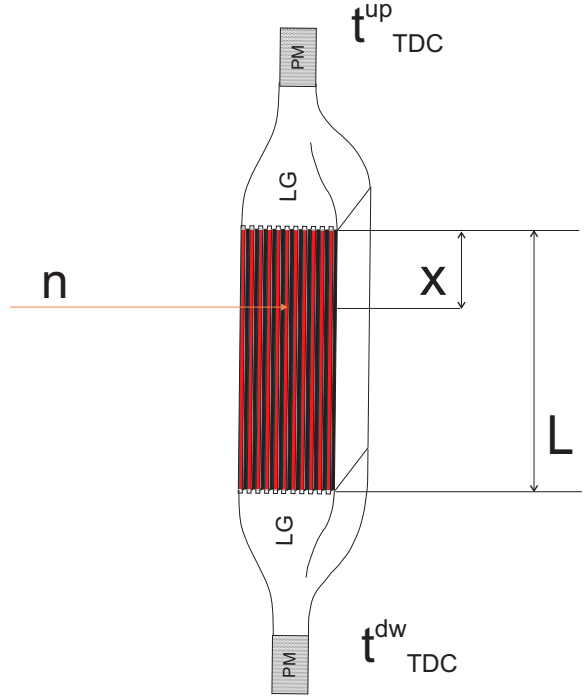,width=0.75\textwidth,angle=0.}}}
\vspace{-3.0cm}
{\caption{   Definition of variables $x$ and $L$ used in the text.
             The scheme of the detection module is the same as in figure 3.3.
            }}
\end{figure}

This can be expressed as:

\begin{equation}
     {T}^{up}_{TDC} = {t}_{real} + \mbox{offset}^{up} + {{x} \over c_L} - T_{trigger},
\end{equation}

\begin{equation}
     {T}^{dw}_{TDC} = {t}_{real} + \mbox{offset}^{dw} + {{L-x} \over c_L} - T_{trigger},
\end{equation}
where $L$ stands for the length of a single module, $x$ denotes the distance between the
upper edge of the active part of the detector and the point at which a neutron induced the hadronic reaction,
$t_{real}$ is the time at which the scintillator light was produced,  $T_{trigger}$
represents the time at which the trigger signal arrives at the TDC converter, and c$_L$ denotes the velocity of the light
signal propagation inside the scintillator plates. The parameters offset$^{up}$
and offset$^{dw}$ denote the time of propagation of signals from the upper and lower edge
of the scintillator to the TDC unit.
\par
Applying  equations 4.1, 4.2, and 4.3 one can calculate a relation between $T^{exp}$ and $t_{real}$:
\begin{equation}
   T^{exp} = t_{real} + {{\mbox{offset}^{up} + \mbox{offset}^{dw} + {{L} \over c_L} } \over 2} - T_{trigger}
           = t_{real} + \mbox{offset} - T_{trigger}
\end{equation}
The value of ``offset'' comprises all delays due to the utilized electronic circuits,
and it needs to be established for each segment. It is worth to note, that this
time of the neutron detector signal is independent
of the hit position,  as it can be deduced from equation 4.4 and was proven experimentally,
and depends on the
difference between the time of light generation and the trigger only.

\subsection{Relative timing between modules}

Instead of determining the value of ``offset'' from equation 4.4 for each detection unit
separately, the relative timing between modules will be first established and then the
general time offset connecting the timing of all segments with the other detectors of the COSY--11
setup will be found.
In order to establish relative time offsets for all
single detection units, distributions of time differences
between neighbouring modules were derived from experimental data.
A time difference measured between two modules can be expressed as:
\begin{equation}
   \Delta_{ij} = T^{exp}_{i} - T^{exp}_{j} =
    {t}_{i}^{real} - {t}_{j}^{real} + (\mbox{offset}_{i} - \mbox{offset}_{j}),
\end{equation}
where $T^{exp}_{i}$ and $T^{exp}_{j}$ stand for the time registered by the $i^{th}$
and $j^{th}$ module, respectively. Example of ${\Delta}_{ij}$ spectra determined from a measurement
carried out in January 2003 with a hydrogen target and a deuteron beam accelerated to the momentum
of $p_b$ = 3.204 GeV/c are presented in figure 4.2. The time differences were calculated assuming
that all constants (offset) are equal to zero (see eq. 4.5).
One can note that the peaks are shifted and additionally the distributions contain long tails.
The tails reflect the velocity distribution of the secondary particles.
Corresponding spectra of time differences between the modules,  shown
in the figure 4.3, were generated using a
GEANT--3 code. To produce these spectra the quasi--free $dp~\to~ppn_{sp}$ reaction has been simulated.
The details are described in the appendix A. The values of the relative time offsets
were determined using a dedicated program written in the Fortran 90 language~\cite{rozek-raport,rozek}.
It adjusts values of offsets such that time differences obtained from experimental data
and from simulation equals to each other for each  pair of detection units.
Furthermore, from the width of the spectra one can receive the information
about the time resolution of
a single module, which was found to be 0.4 ns~\cite{rozek-raport}.

\begin{figure}[H]
\parbox{0.3\textwidth}{\epsfig{file=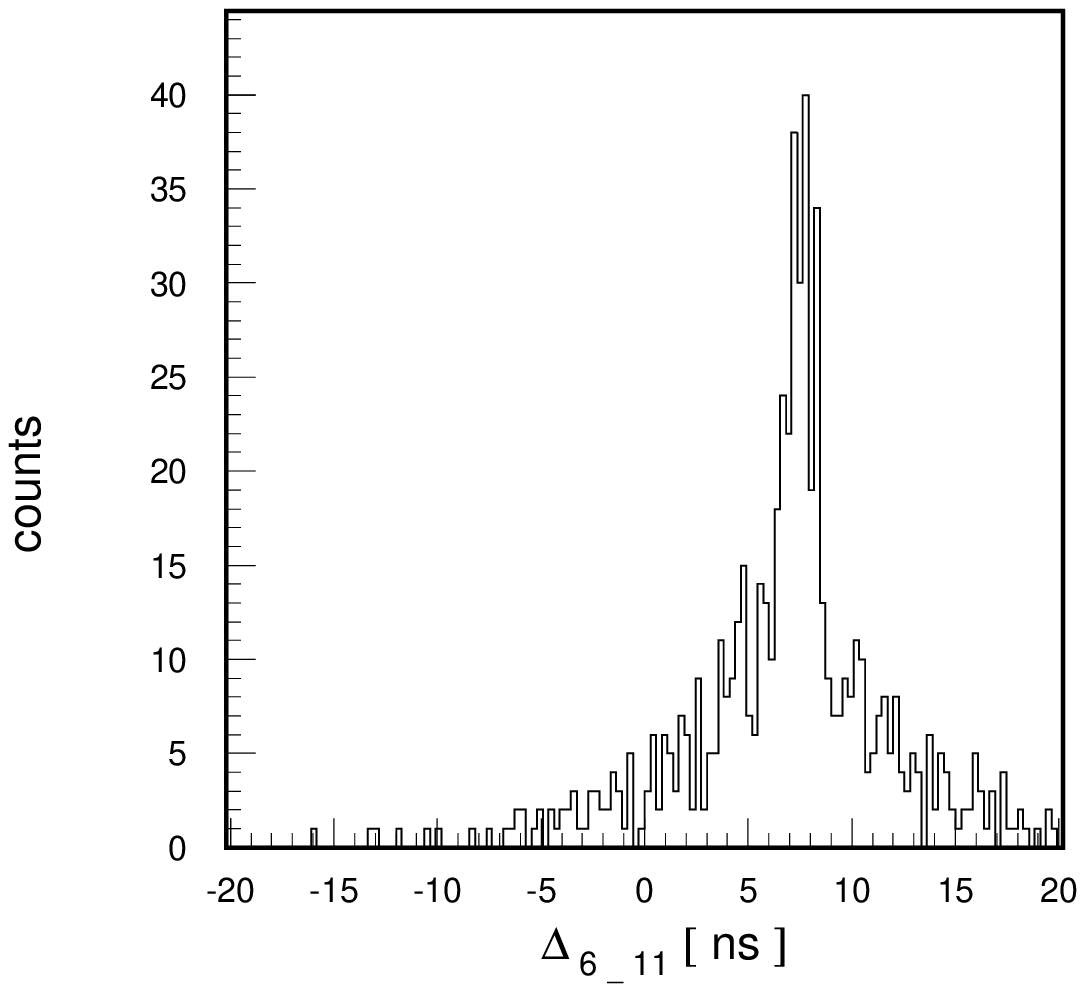,width=0.4\textwidth,angle=0.}}
\parbox{0.3\textwidth}{\epsfig{file=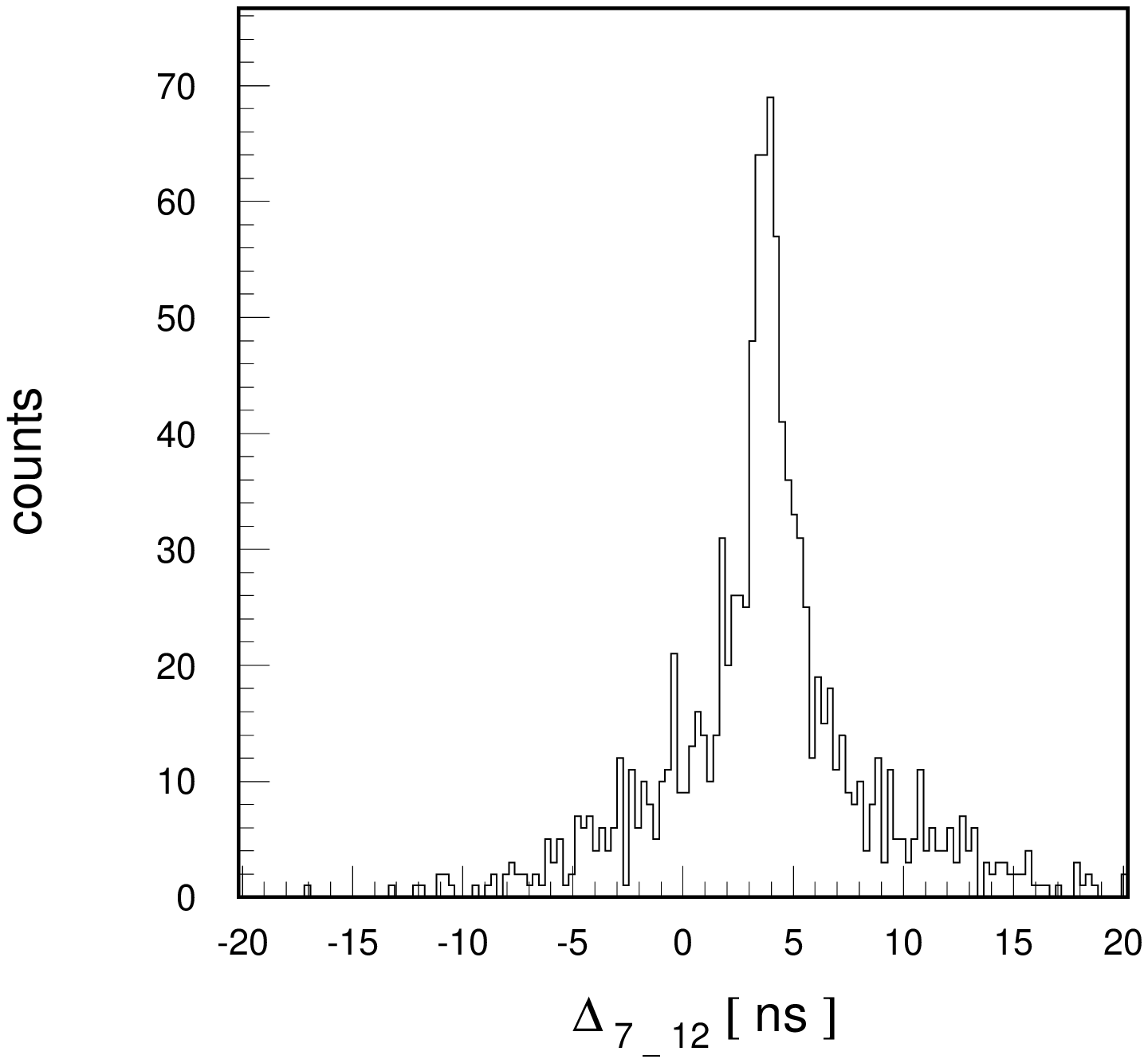,width=0.4\textwidth,angle=0.}}
\parbox{0.3\textwidth}{\epsfig{file=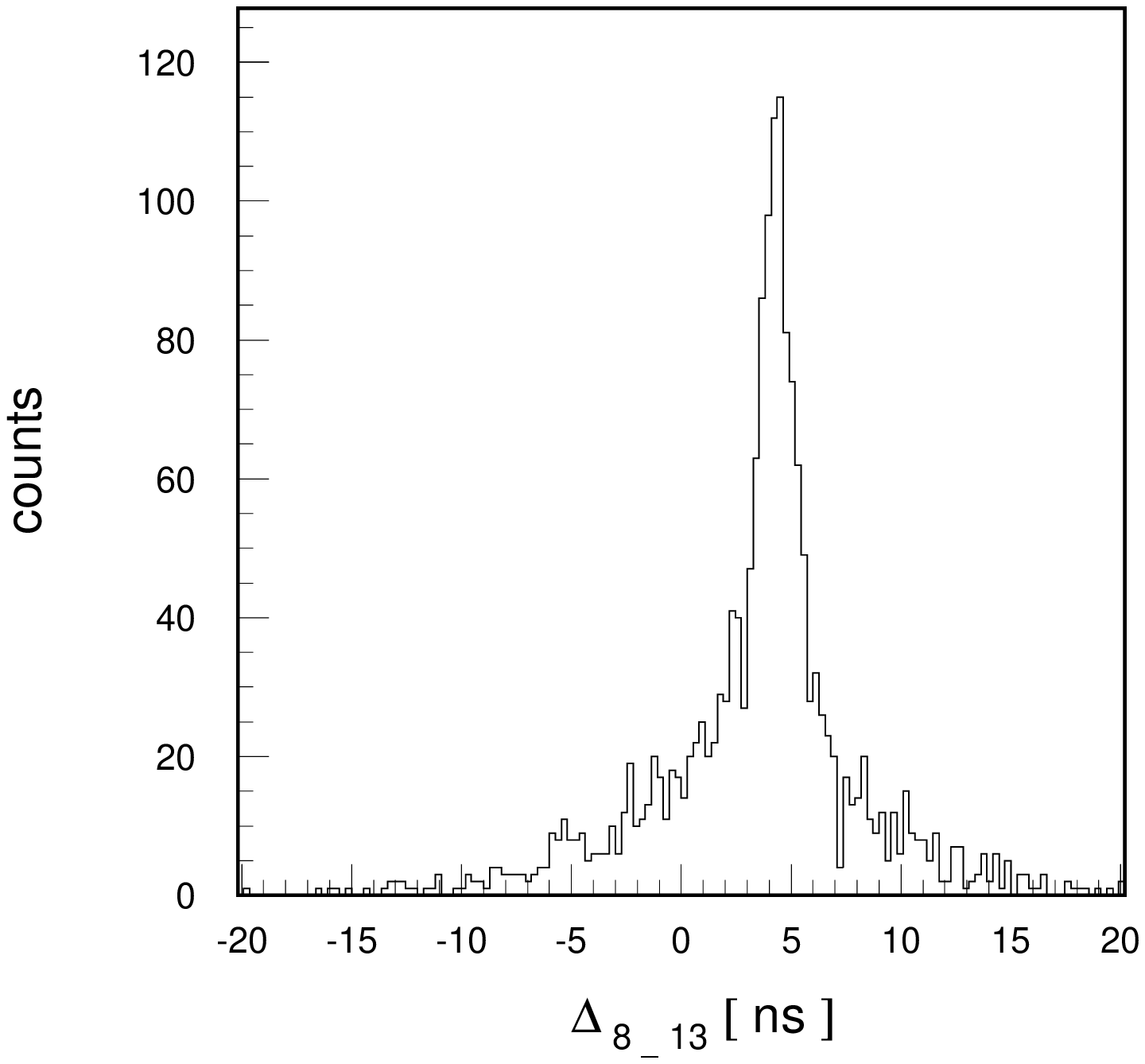,width=0.4\textwidth,angle=0.}}

\vspace{-0.5cm}

{\caption{
             Distribution of the time difference between the $6^{th}$ and the $11^{th}$,
             the $7^{th}$ and the $12^{th}$, and the $8^{th}$ and the $13^{th}$ module of
             the neutron detector, as determined before the calibration.}}
\end{figure}

\vspace{-0.5cm}

\begin{figure}[H]
\parbox{0.3\textwidth}{\epsfig{file=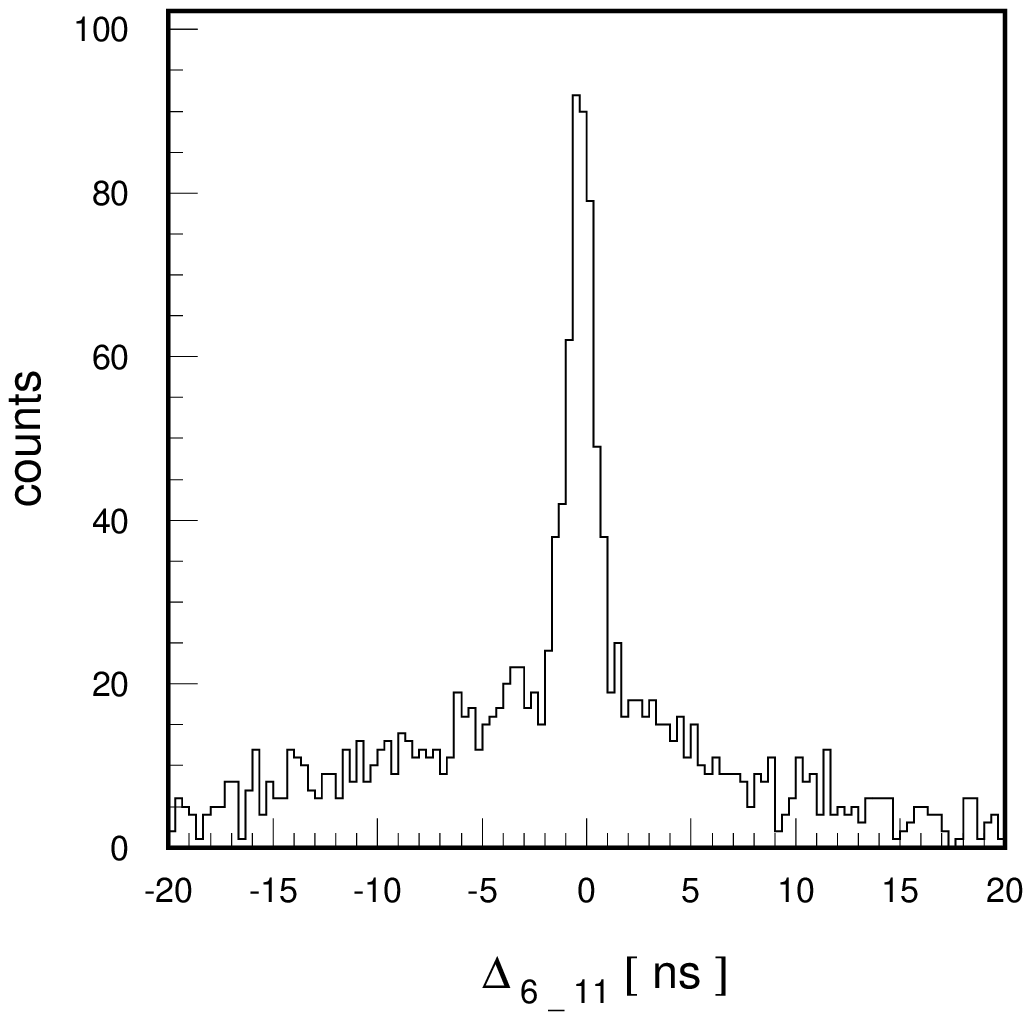,width=0.4\textwidth,angle=0.}}
\parbox{0.3\textwidth}{\epsfig{file=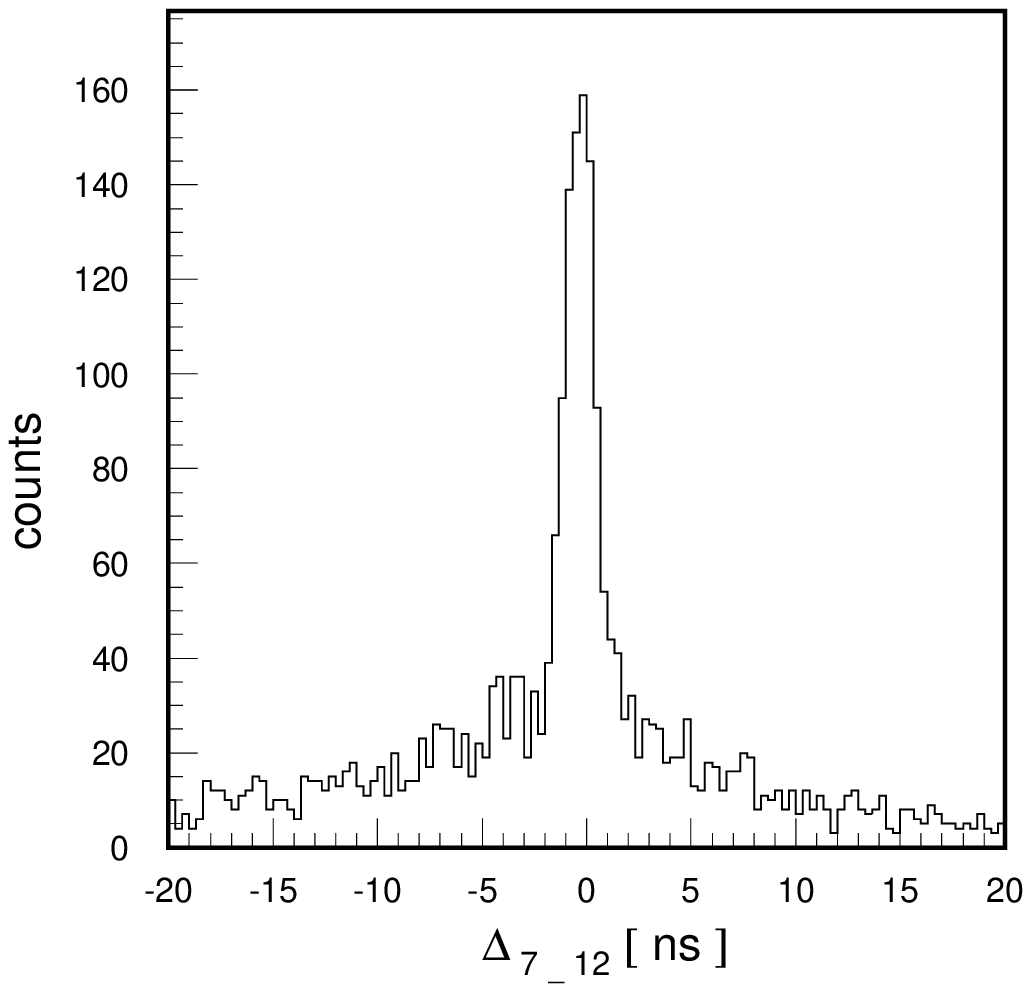,width=0.4\textwidth,angle=0.}}
\parbox{0.3\textwidth}{\epsfig{file=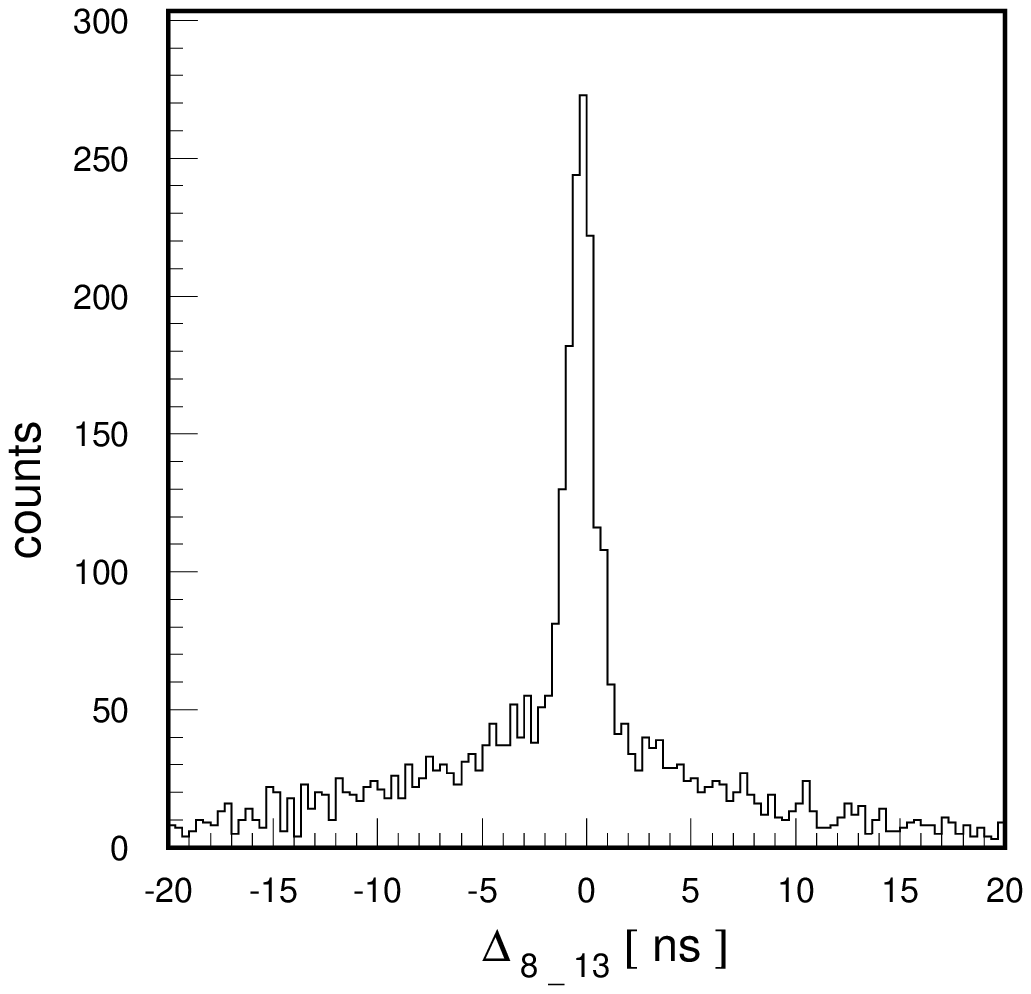,width=0.4\textwidth,angle=0.}}

\vspace{-0.5cm}

  {\caption{
             Simulated distribution of the time difference between the $6^{th}$ and the $11^{th}$,
             the $7^{th}$ and the $12^{th}$, and the $8^{th}$ and the $13^{th}$ module of
             the neutron detector.}}
\end{figure}


\vspace{-0.5cm}

\begin{figure}[H]
\parbox{0.3\textwidth}{\epsfig{file=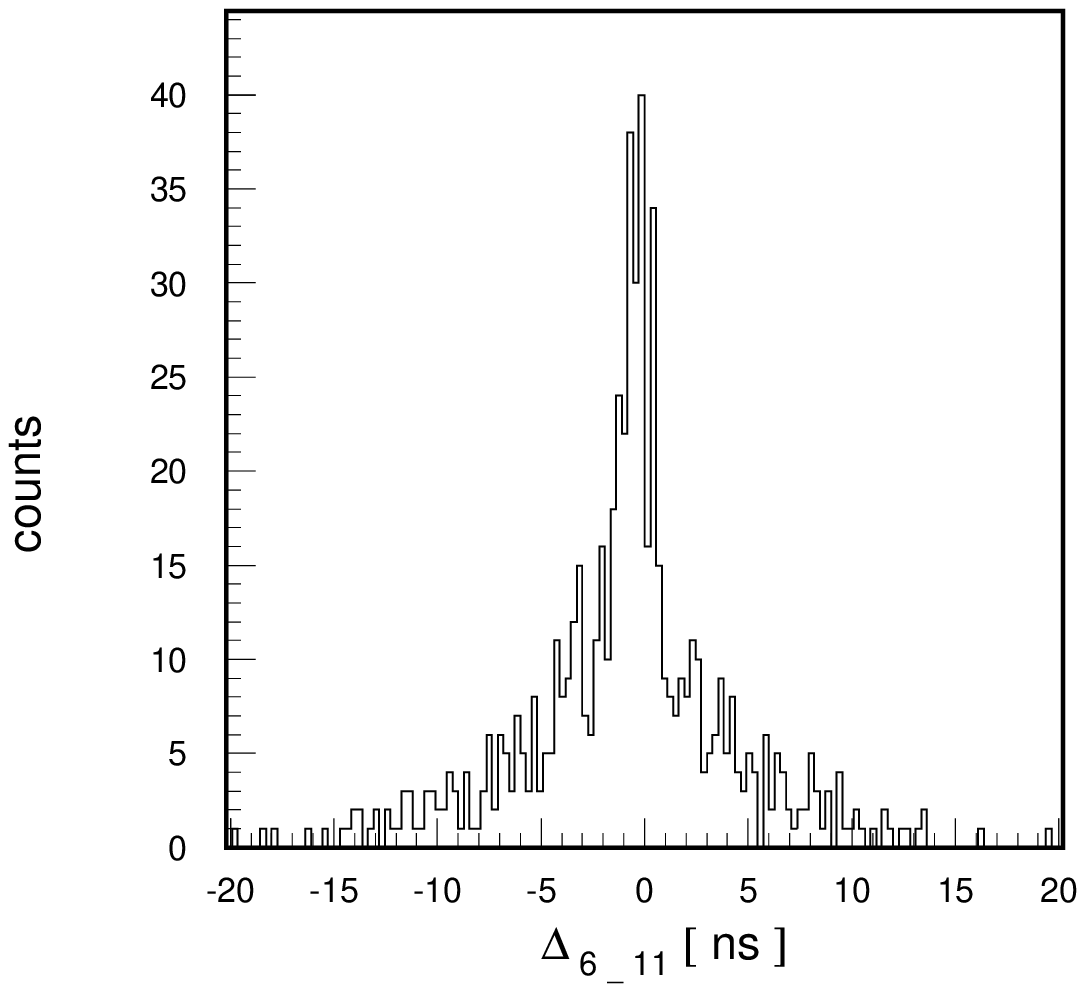,width=0.4\textwidth,angle=0.}}
\parbox{0.3\textwidth}{\epsfig{file=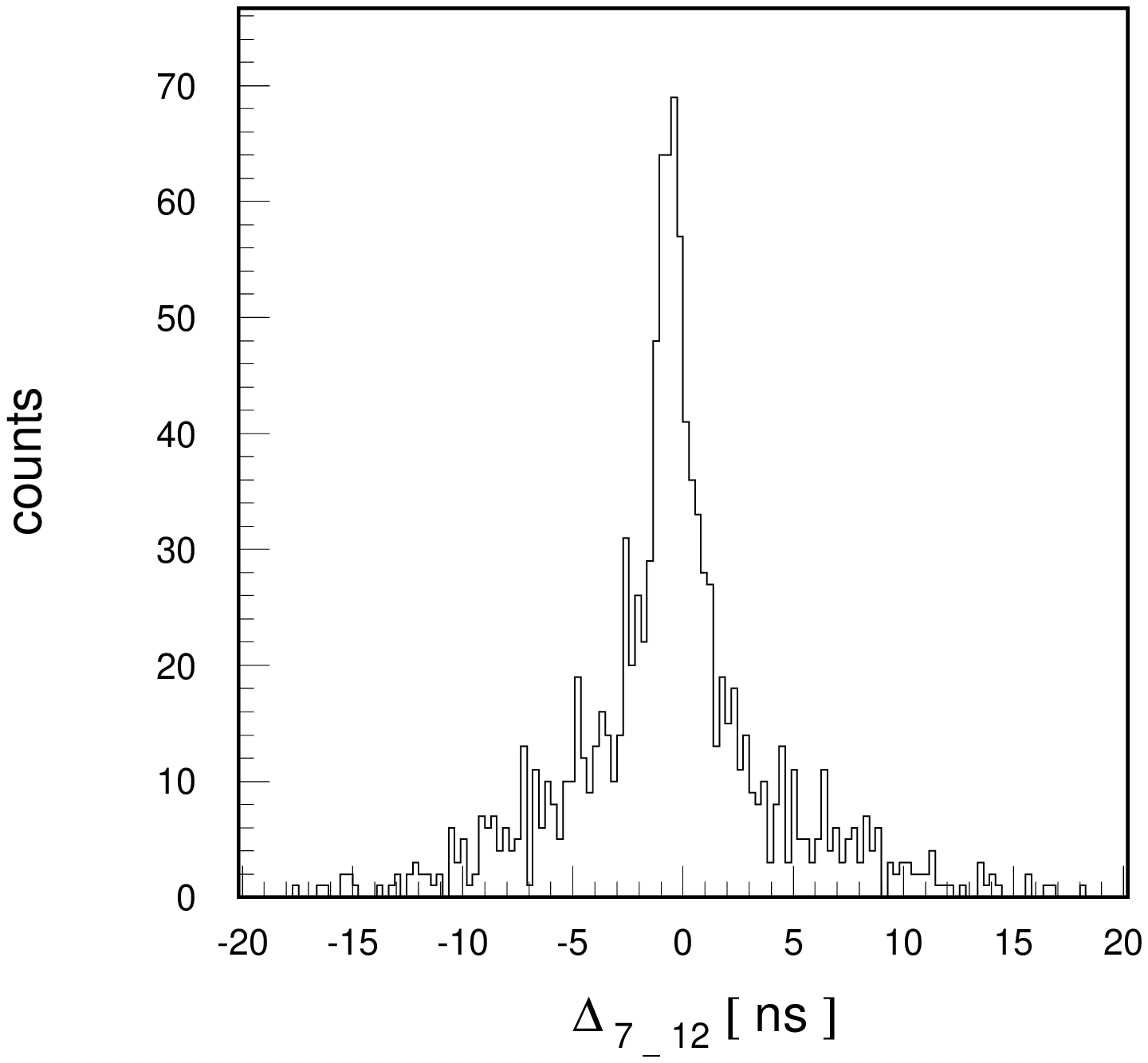,width=0.4\textwidth,angle=0.}}
\parbox{0.3\textwidth}{\epsfig{file=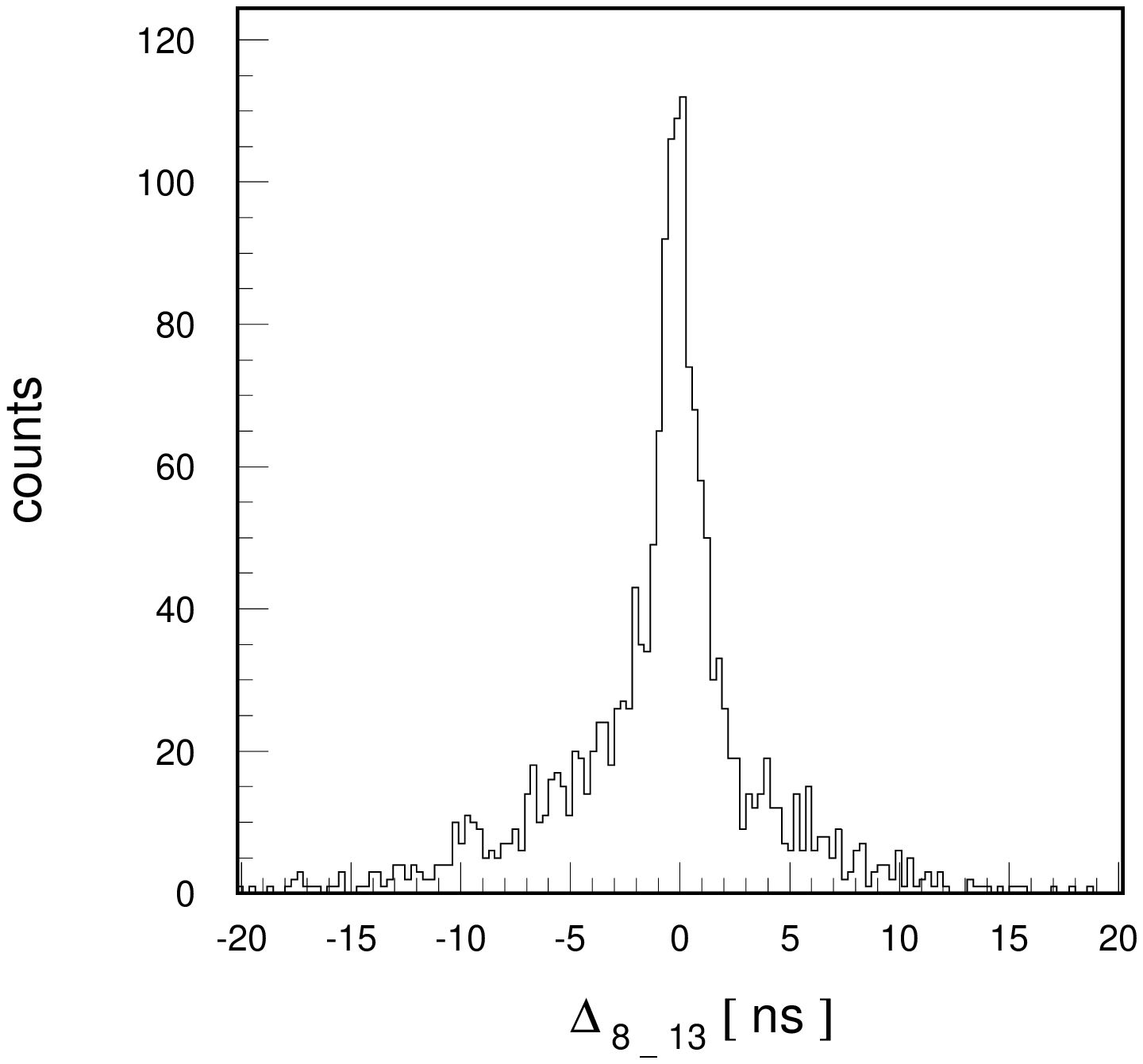,width=0.4\textwidth,angle=0.}}
\parbox{0.99\textwidth}

\vspace{-0.5cm}

  {\caption{
             Distribution of the time difference between the $6^{th}$ and the $11^{th}$,
             the $7^{th}$ and the $12^{th}$, and the $8^{th}$ and the $13^{th}$ module of
             the neutron detector, as obtained after the calibration.}}
\end{figure}
Figure 4.4 presents  experimental distributions of time differences between
neighbouring modules as determined after the calibration.
Examining figure 4.3 and 4.4 it is evident that now the peaks are positioned precisely at
the value expected from simulation.

\vspace{-0.55cm}

\subsection{General time offset}

For the determination of the momentum of neutrons eg. in the analysis of reactions
like $dp~\to~ppn\gamma$, the time between the reaction moment
and the hit time in the neutron detector has to be determined for each event.
The time of the reaction can be deduced from the time when the charged particle
crosses the S1 detector (see fig. 3.1) assuming that this particles trajectory and
velocity can be reconstructed.
To perform the calculation of the  time--of--flight between target and neutron counter
a general time offset of the neutron detector with respect to the
S1 detector has to be established. For this purpose the quasi--free $dp~\to~ppn_{sp}$
reaction will be used. In this type of reactions the proton bound in a deuteron scatters elastically
on a target proton, whereas the neutron considered as a spectator does not interact with
the proton, but escapes untouched and hits the neutron detector.
\par
Data have been taken at a beam momentum of 3.204 GeV/c close to thereshold of the
$dp~\to~dp\eta$ process. Events corresponding to the $dp~\to~ppn_{sp}$ reaction
have been identified by measuring the outgoing charged as well as neutral ejectiles.

\begin{figure}[H]
\centerline{\parbox{0.6\textwidth}{\epsfig{file=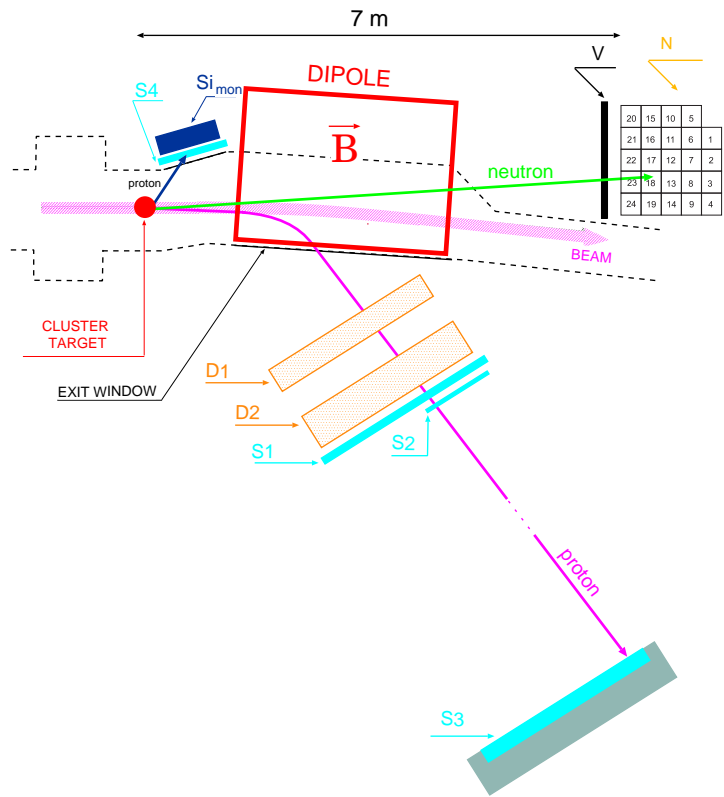,width=0.65\textwidth,angle=0.}}}
\parbox{0.99\textwidth}
  {\caption{  Schematic view of the COSY--11 detection system.
              Only detectors used for the measurement of
              elastic scattered protons and spectator neutrons are shown.
            }}
\end{figure}

Fast protons are detected by means of the drift chambers (D) and scintillator hodoscopes (S1-S3).
Protons scattered under large angles are measured in a position sensitive silicon detector $(Si_{mon})$.
Neutrons are registered in the scintillator--lead sandwich detector (N).
\par
The time-of-flight between the target and the neutron detector $({TOF}^{neut})$
is calculated as a difference between the time of the module
in the neutron detector which fired as the first one ($t_{n}^{real}$)
and the time of the reaction $(t^{real}_{r})$.
\begin{equation}
    {TOF}^{neut} =  t_{n}^{real} - t^{real}_{r}
\end{equation}
Time of the reaction is obtained from backtracking of protons through the known
magnetic field and the time measured by the S1 detector ($T^{S1}$). Thus $t^{real}_{r}$
can be expressed as:
\begin{equation}
    t^{real}_{r} = t_{S1} - TOF = T^{S1} - \mbox{offset}^{S1} + T_{trigger} - TOF ,
\end{equation}
where $TOF$ denotes the time--of--flight between the S1 counter and the target,
and offset$^{S1}$ denotes all delays of signals from the S1 detector.
Since time in the neutron detector reads:
\begin{equation}
    t_{n}^{real} = T_{n} - \mbox{offset}^{N} + T_{trigger}
\end{equation}
we have:
\begin{equation}
   {TOF}^{neut} = T_{n} + TOF - T^{S1} - \mbox{offset}^{N} + \mbox{offset}^{S1}
      = T_{n} + TOF - T^{S1} + \mbox{offset}^{G}
\end{equation}
 By offset$^{G}$ the general time offset of the neutron detector in respect to S1 is denoted.
In order to establish this global time offset,
the time--of--flight spectrum derived from experimental data for the $dp~\to~ppn_{sp}$ reaction
was compared with the corresponding distribution which was reconstructed from the signals
simulated in the detectors (see figure 4.6 left).
To arrive at the same statistic as was achieved in experiment, $8 \cdot 10^{6}$ events for the
$dp~\to~ppn_{sp}$ reaction were simulated using a GEANT--3 code. In the simulation
the time resolution
of a single module of $\sigma~=~0.4$~ns was taken into account.

\begin{figure}[H]
\parbox{0.45\textwidth}{\hspace{-0.5cm}\epsfig{file=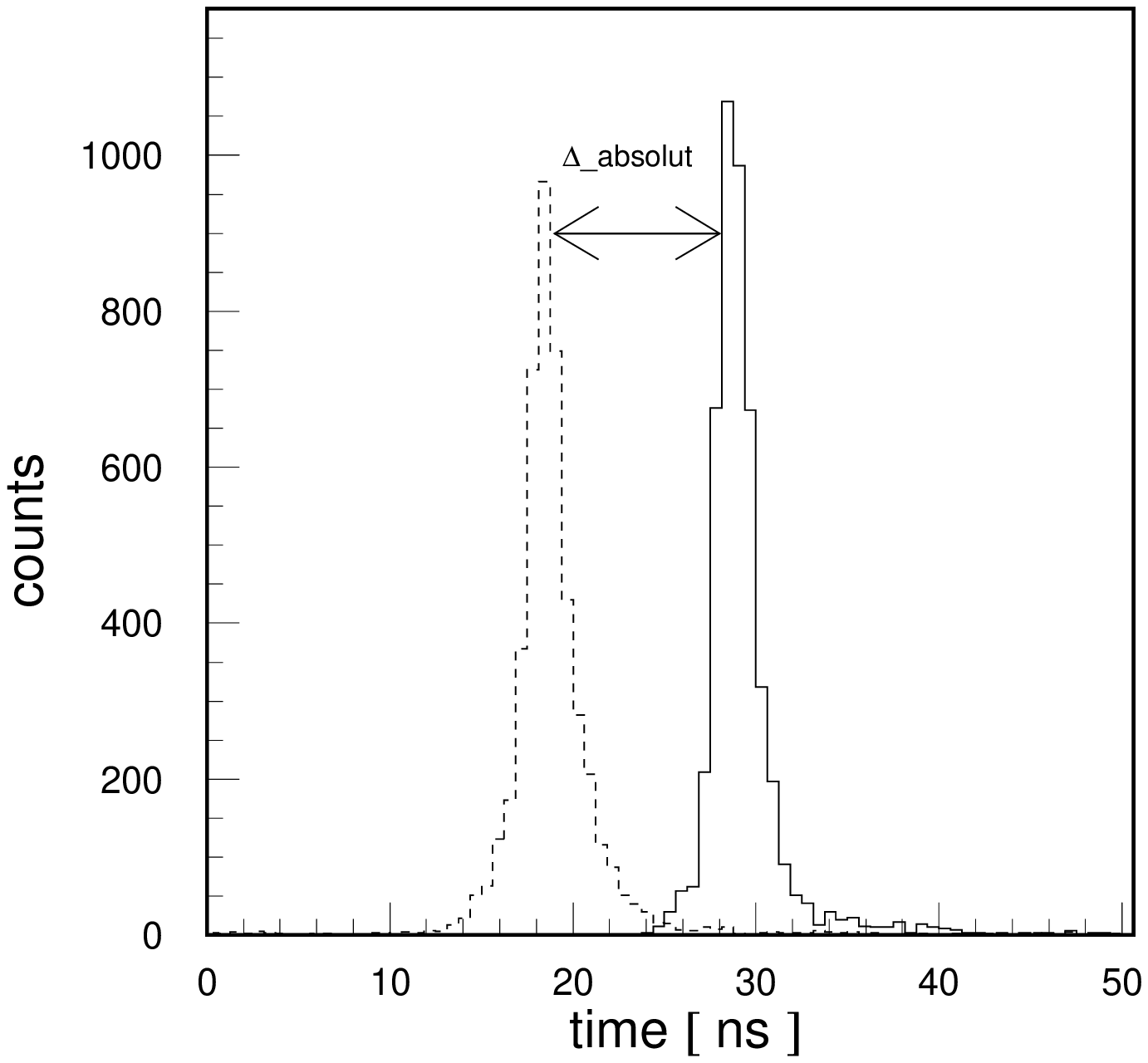,width=0.590\textwidth,angle=0.}}
\parbox{0.45\textwidth}{\hspace{-1.5cm}\epsfig{file=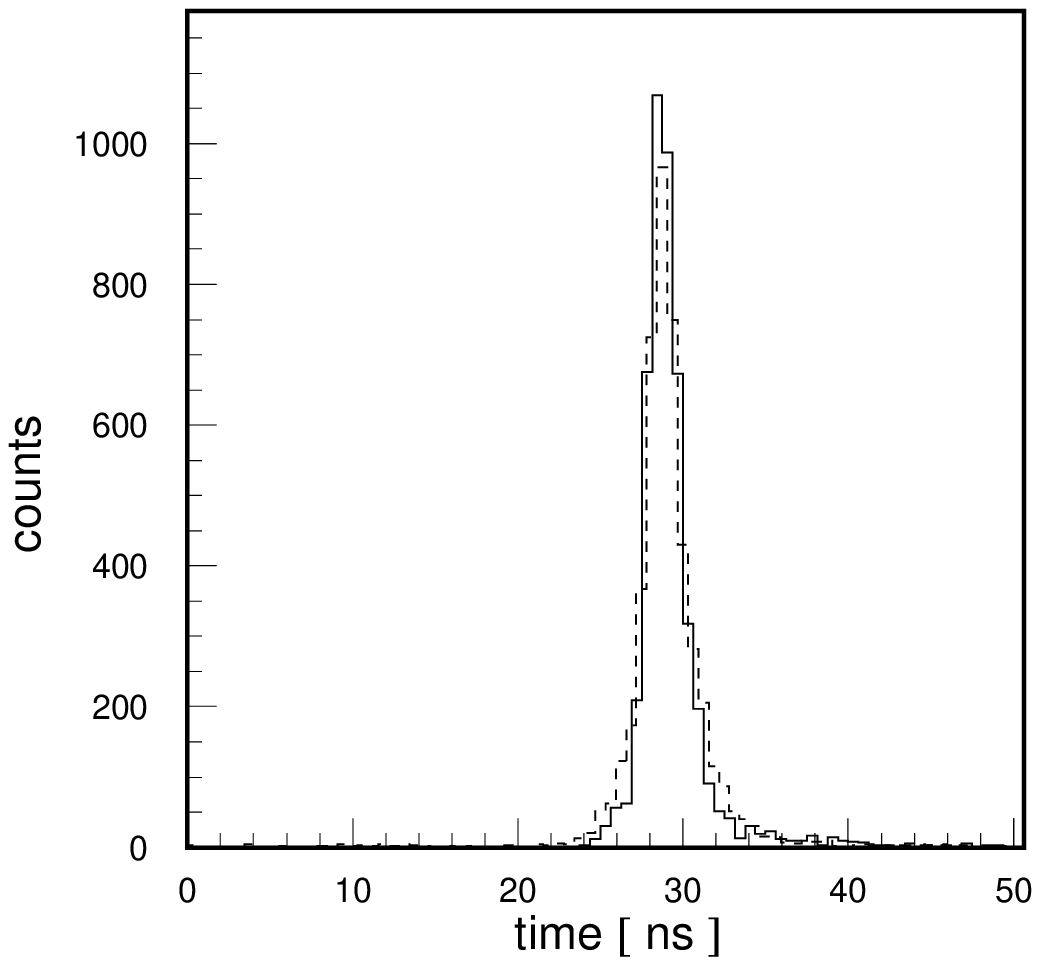,width=0.590\textwidth,angle=0.}}
       {\caption{ {\bf (left)}  Time--of--flight distribution between the target
                    and the neutron detector as obtained before the general time offset
                    had been determined. The dashed histogram denotes the experimental data
                    whereas the solid one depicts the result of the simulation.
                 {\bf (right)} Experimental time-of-flight distribution (dashed histogram)
                    calculated between the target
                    and the neutron detector with general time offset equal to 13.6 ns
                    compared to the Monte--Carlo simulation (solid histogram). The latter was
                    normalized in number of counts to the experimental data.
                   }}
\end{figure}
The simulated spectrum was normalized so that the integrals of both distributions are equal.
With a general time offset of 13.6~ns the experimental distribution corresponds to the
simulated one as shown in figure 4.6 (right). The main
cause of the smearing of the considered time distribution is the Fermi momentum
of the nucleon inside the deuteron. The time resolution
and dimensions of the detector are of minor importance.

\section{ Momentum resolution }
The experimental resolution of the missing mass determination eg. in the analysis of reactions
like $pn~\to~pn\eta'$ strongly rely on the accurate measurement of the momentum of the neutrons~\cite{moskal-hadron}.
Therefore, the momentum resolution of the neutron detector has to be elaborated.
After the neutron and gamma quanta are identified the
momentum of neutrons is calculated from the time--of--flight between the
target and the neutron detector according to the formula 3.1.
\par
Monte Carlo studies of the $dp~\to~ppn_{sp}$ reaction have been performed in order
to establish the momentum resolution of the neutron detector. Figure 4.7 presents the difference
between the generated neutron momentum ($P_{gen}$)
and the reconstructed neutron momentum
from signals simulated in the detectors ($P_{rec}$).
\begin{equation}
   \Delta P = P_{gen} - P_{rec}
\end{equation}
The value of ($P_{rec}$) was calculated
taking into account the time resolution of the neutron detector ($\sigma$~=~0.4~ns)
as well as the time resolution of the S1 counter ($\sigma$~=~0.25~ns).
\begin{figure}[H]
\centerline{\parbox{0.6\textwidth}{\epsfig{file=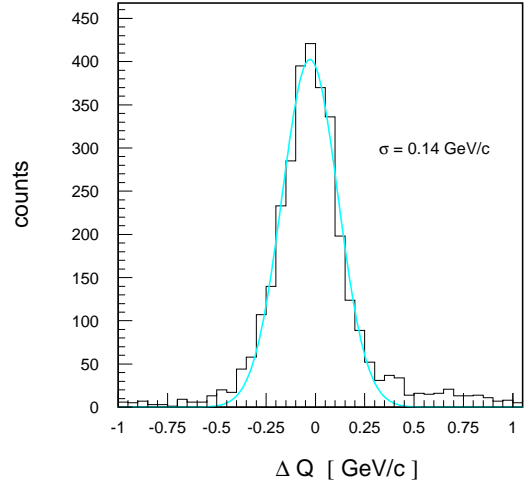,width=0.65\textwidth,angle=0.}}}
\parbox{0.99\textwidth}
      {\caption{
                 Difference between generated ($P_{gen}$)
                 and reconstructed ($P_{rec}$) neutron momenta for the
                 $dp~\to~ppn_{sp}$ reaction simulated at deuteron beam momentum
                 of 3.204~GeV/c.
      }}
\end{figure}
The distribution of $\Delta P$ was fitted by a Gaussian function resulting in a momentum
resolution of $\sigma$(P)~=~0.14~GeV/c. Consequently the
fractional momentum resolution $\sigma$(P)/P for neutrons with
momentum value of 1.6~GeV/c is equal to 8{\%}.
This fractional resolution changes with the momentum of the neutron (see Appendix B) and eg.
for the neutrons produced at threshold for the $pn~\to~pn\eta$ reaction it amounts to 3$\%$.

\chapter{ Analysis of the experimental data}

\markboth{\bf Analysis of the experimental data}
         {\bf Analysis of the experimental data}

For the first time at the COSY--11 experiments  signals from  $\gamma$--quanta
were observed in the
time--of--fligth distribution (for the neutral particles) measured between the target and
the neutral particle detector. A corresponding spectrum is shown in figure 5.1(left). The data are from
an experiment carried out using a deuteron target and
a proton beam with a momentum of 2.075 GeV/c~\cite{moskal-hadron}.
In addition to a broad distribution originating from neutrons, a sharp
peak from ${\gamma}$ rays is seen at a value of about 25 ns.
A Monte Carlo simulation performed for the $pn~\to~pn\eta$ reaction,
which is one of the possible processes contributing to the neutron time-of-flight
distribution, is shown in figure 5.1(right)\\
\begin{figure}[H]
\parbox{0.45\textwidth}{\hspace{-0.5cm} \epsfig{file=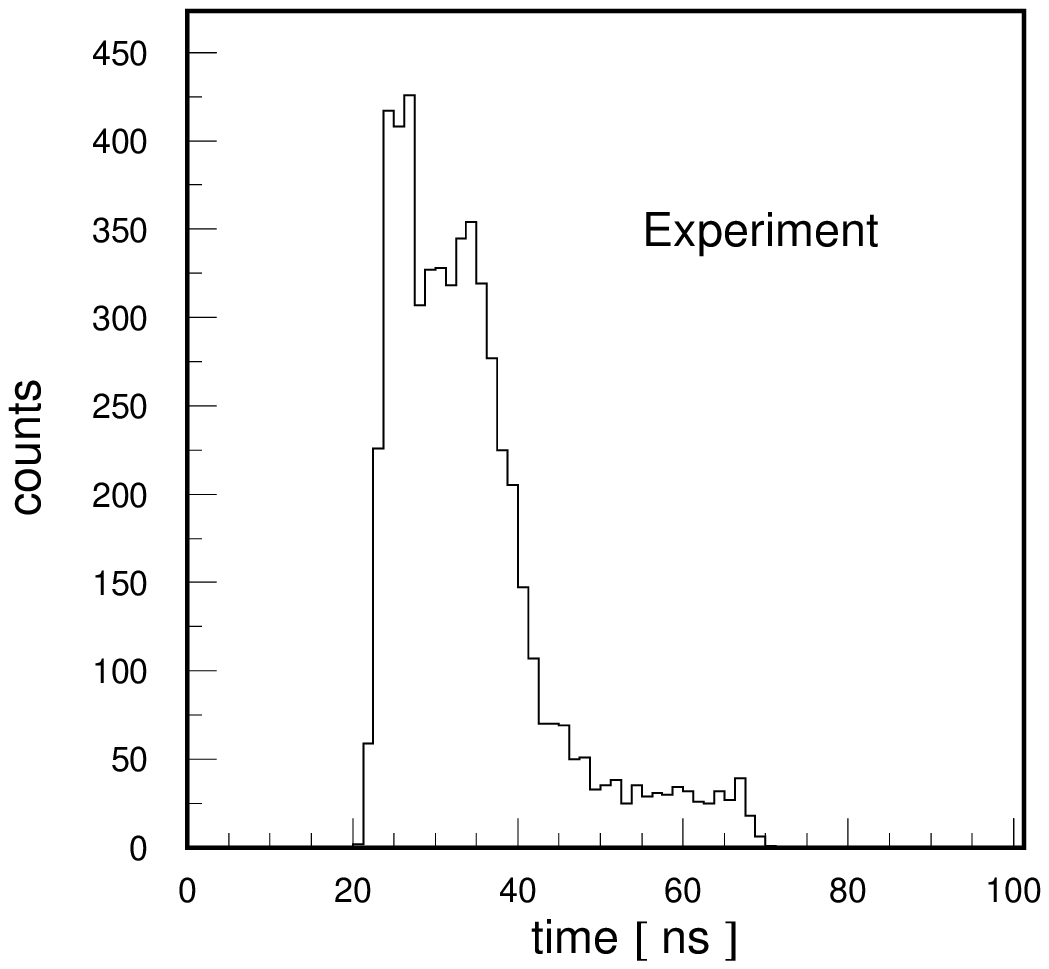,width=0.59\textwidth,angle=0.}}
\parbox{0.45\textwidth}{ \hspace{-1.5cm} \epsfig{file=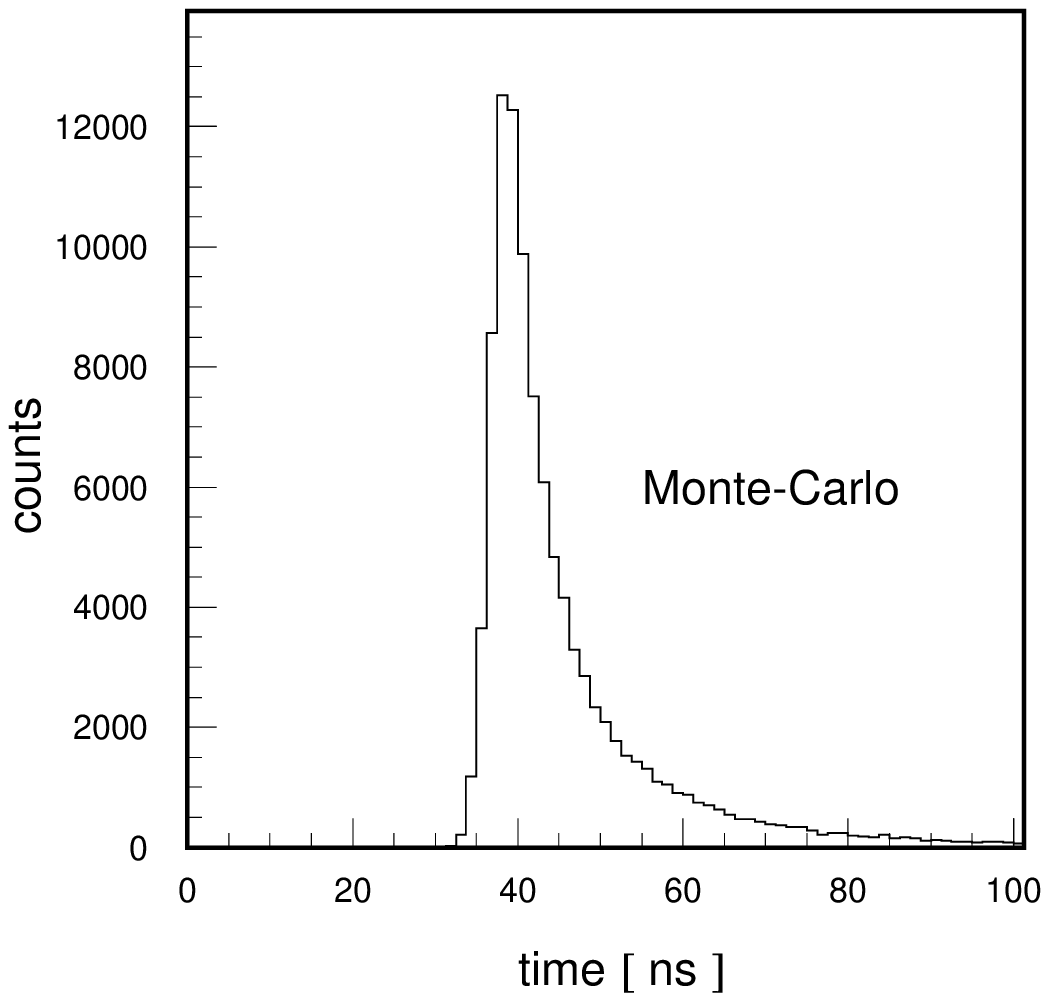,width=0.59\textwidth,angle=0.}}
      {\caption{  Time--of--flight distribution determined between
                  the target and the neutron detector.\\
                  {\bf (left)} Experimental
                  spectra, and  {\bf (right)} Monte Carlo simulation.
                  Figure adapted from~\cite{przerwa}
                        }}
\end{figure}

 In this chapter results of the analysis aiming for the identification of bremsstrahlung
radiation in the data taken in January 2003 will be presented. The measurement
 was carried out with a deuteron beam and
hydrogen cluster target at the cooler synchrotron COSY--J{\"u}lich by means
of the COSY--11 detection system, and its primordial aim was to investigate the
$dp~\to~dp\eta$ reaction close to the threshold~\cite{smyrski}. The experiment was performed at four different
deuteron beam momenta between $p_{beam}$ = 3.165 and 3.204 GeV/c. During the run
with a beam momentum of $p_{beam}$ = 3.204 GeV/c, an additional trigger with
neutron detector --- referred to T8 --- was
set up for the registration of charged ejectiles in coincidences with neutrons or gamma quanta.
These conditions can be written symbolically as:
\vspace{-0.1cm}
$$ T8 = {S1}_{\mu=2} \wedge N_{\mu \geq 2} \wedge S3_{\mu=1}, $$
which means that
two signals in the S1 detector, one signal in the S3 detector and, at least two signals
in the neutron detector were demanded.
\par
Possible reactions with gamma quanta in the final state can be divided into
two groups, viz: free $dp~\to~dp\gamma$ and $dp~\to~^{3}He\gamma$ reactions,
and quasi--free $dp~\to~dp_{sp}\gamma$, $dp~\to~pp\gamma n_{sp}$, $dp~\to~pn\gamma p_{sp}$
reactions. In case of quasi--free reactions one of the nucleons bound in a beam deuteron
is treated as a spectator ($p_{sp}, n_{sp}$) and does not take part in the reaction.
\par
As a first step  of the data analysis events with simultaneous signals in any of the
drift chambers and the neutron detector were selected.
Figure 5.2 presents the time--of--flight distributions
between the target and the neutron detector obtained assuming that in coincidence with a neutral particle
also a proton (left) or deuteron (right) was identified based on signals from drift chambers
and scintillator hodoscopes. The signals from $\gamma$--quanta do not appear in the distributions
which are
predominantly due to quasi--free elastic
$dp~\to~ppn_{sp}$ and $dp~\to~dn\pi^{+}$ reactions.
\begin{figure}[H]
\parbox{0.40\textwidth}{\hspace{-0.5cm}\epsfig{file=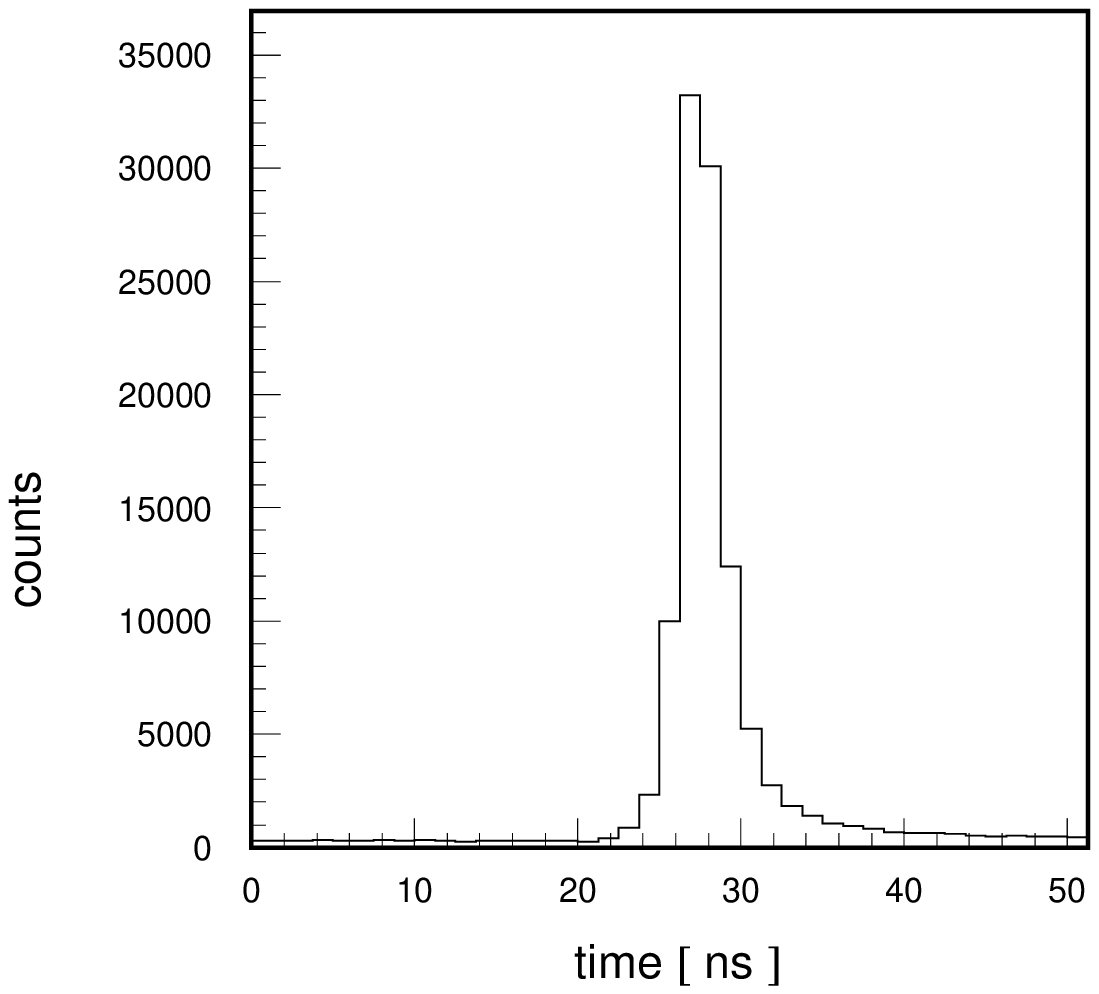,width=0.5\textwidth,angle=0.}}
\parbox{0.40\textwidth}{ \hspace{-1.0cm} \epsfig{file=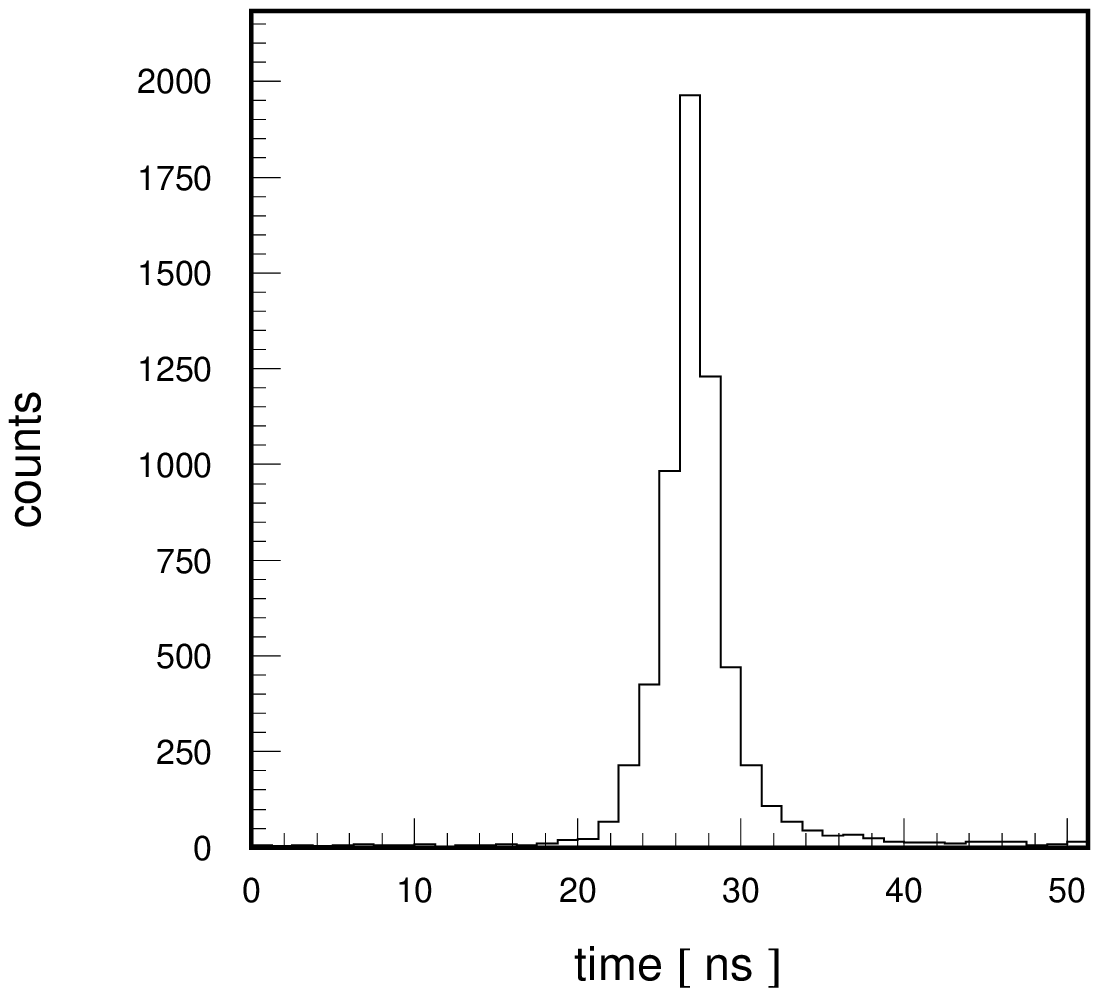,width=0.5\textwidth,angle=0.}}
       {\caption{  Experimental time--of--flight distributions determined between
                   the target and the neutron detector obtained under the condition
                   that additionally one proton  {\bf (left)} or one deuteron
                  {\bf (right)} was registered in drift chambers.
                        }}
\end{figure}
The charged ejectiles can be well identified as shown in the figure 5.3.
Three clear peaks evidently visible in this figure correspond to the squared
mass of pion, proton, and deuteron.
The mass of the particle is calculated from its momentum~--- reconstructed from
tracking back through magnetic field to the target point~---
and velocity determined from the time--of--flight measured between S1 and S3 detector.

\begin{figure}[H]
\centerline{\parbox{0.75\textwidth}{\epsfig{file=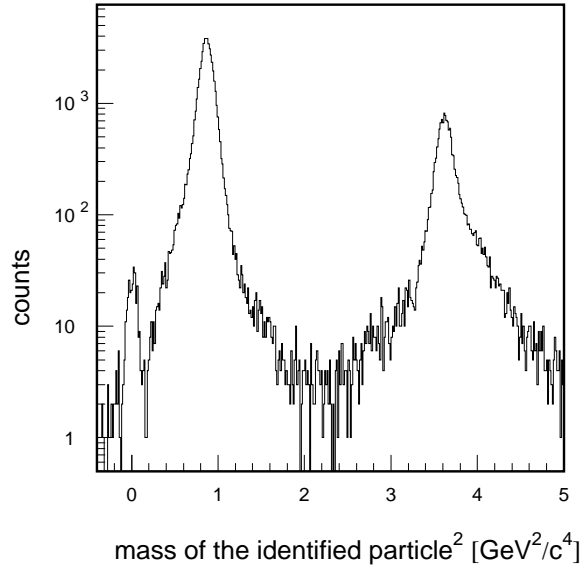,width=0.75\textwidth,angle=0.}}}
      {\caption{   Distribution of the squared mass of charged particles
                   originating from the deuteron--proton reaction performed
                   at a beam momentum of 3.204~GeV/c.
               }}
\end{figure}

\section{Identification of the $dp~\to~dp\gamma$ reaction}
 In order to identify the $dp~\to~dp\gamma$ reaction events with two tracks in drift chambers
and a simultaneous signal in a neutron detector have been selected. In figure 5.4 squared mass of
one particle is plotted versus squared mass of the other registered particle.
Base on this figure measured reactions can be grouped according to the type of ejectiles.
\begin{figure}[H]
\centerline{\parbox{0.7\textwidth}{\epsfig{file=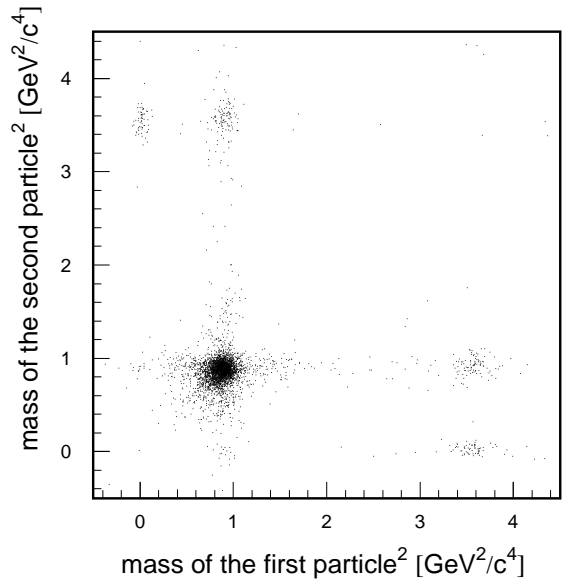,width=0.75\textwidth,angle=0.}}}
       {\caption{  Scatter plot of invariant masses determined for events with
                   two charged particles measured in coincidence.
               }}
\end{figure}
\vspace{0.5cm}
Thus reactions with two protons, proton and pion, proton and deuteron,
and pion and deuteron can be cleary separated.
Figure 5.5 shows the experimental distribution of
the time--of--flight between the target and neutral--particle detector with the requirement
that two charged particles were registered and that one of them was identified as a proton and
the other as a deuteron.
In this case due to the baryon number conservation, there is only one possible source of a signal
in a neutron detector, namely a gamma quantum. In fact a clear peak around the time of 24.5 ns is visible,
and this is just the value corresponding to the time--of--flight of light on a distance of 7m.
The gamma quanta may originate from the bremsstrahlung reaction ($dp~\to~dp\gamma$)
or from the decay of mesons produced eg. via $dp\to~dp\pi^{0}~\to~dp\gamma\gamma$
or $dp~\to~dp\pi^0\pi^0~\to~dp4\gamma$ reactions. It is possible to distiguish between these
hypothesis calculating the missing mass produced in the $dp~\to~dpX$ reaction.

\begin{figure}[H]
\centerline{\parbox{0.7\textwidth}{\epsfig{file=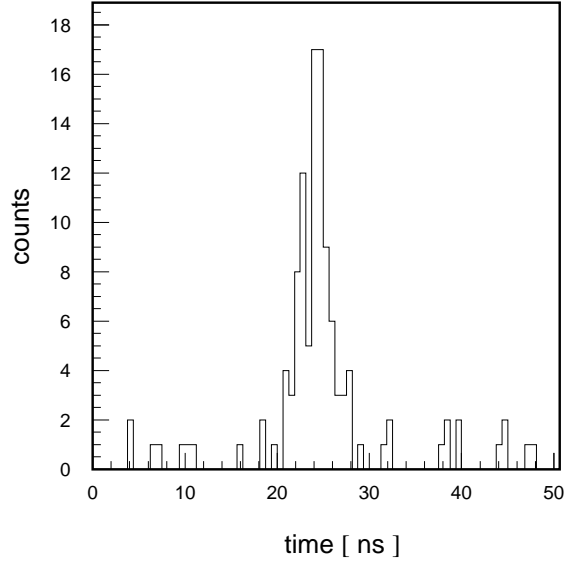,width=0.75\textwidth,angle=0.}}}
      {\caption{  Time--of--flight distribution determined between
                  the target and the neutron detector obtained under the assumption
                  that additionally to a signal in the neutron detector,
                  one proton and  one deuteron were identified from the
                  signals in the drift chambers.
               }}
\end{figure}
\par
Knowing the four momenta of
a proton and a deuteron in the initial and final state, and employing the principle of
 momentum and energy conservation
one can calculate the squared mass of the unmeasured particle or group of particles:
\begin{equation}
 m_{x}^{2} = E_{x}^{2} - {\vec {p_{x}}}^{2} = {(E_{b} + E_{t} - E_{d} - E_{p})}^2 -
  {( \vec p_{b} + \vec p_{t} - \vec p_{d} - \vec p_{p} )}^2
\end{equation}
where,\\
$E_{b}, \vec p_{b}$ is the energy and momentum of deuteron beam,\\
$E_{t}, \vec p_{t}$ is the energy and momentum of proton target,\\
$E_{d}, \vec p_{d}$ is the energy and momentum of outgoing deuteron, and\\
$E_{p}, \vec p_{p}$ is the energy and momentum of outgoing proton.\\

Figure 5.6 shows the squared missing mass distribution as obtained for the
$dp~\to~dpX$ reaction. A significant peak around 0 $MeV^2/c^4$ --- the squared mass of the gamma
quantum --- constitutes an evidence for events associated to the deuteron--proton bremsstrahlung ($dp\gamma$).
In addition  a broad structure at higher masses originating from two pions emitted from the reaction
$dp~\to~dp\pi^0\pi^0$ is visible.
\begin{figure}[H]
\centerline{\parbox{0.7\textwidth}{\epsfig{file=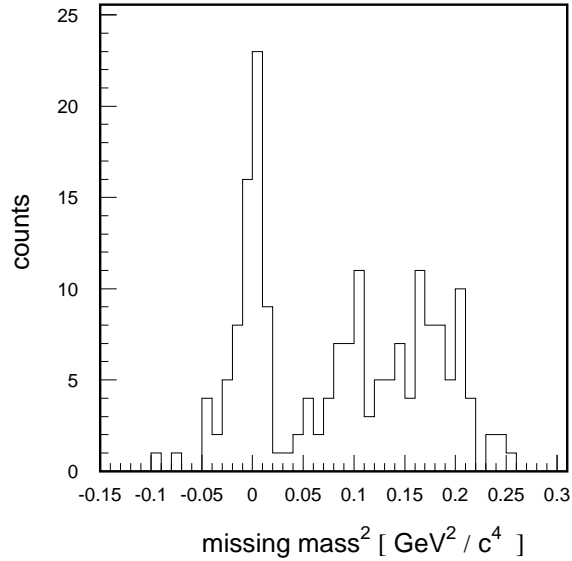,width=0.75\textwidth,angle=0.}}}
      {\caption{ Squared missing mass distribution for
                 the $dp~\to~dpX$ reaction.
               }}
\end{figure}
One of the most important information is the energy spectrum of the gamma quanta.
The energy of gamma quanta can be measured with a very poor accuracy, however
it can be calculated using a simple relation, after selecting events corresponding to the
$dp~\to~dp\gamma$ reaction, namely:
\begin{equation}
  E_{\gamma} = E_b + E_t - E_d - E_p
\end{equation}
where,\\
$E_{b}$ is the energy of deuteron beam,\\
$E_{t}$ is the energy of proton target,\\
$E_{d}$ is the energy of outgoing deuteron, and\\
$E_{p}$ is the energy of outgoing proton.\\

 Figure 5.7 shows the determined spectrum. One sees that the registered gamma quanta
populate predominantly the energy range between 0.8 and 0.9~GeV, which is partially due to
the acceptance of the COSY--11 system which decreases rapidly with increasing kinetic energy
shared by the ejectiles. The detailed conclusions concerning the real energy distribution of the
produced gammas will require careful acceptance corrections which will be performed in the
near future. At present we can consider the observed distribution as an evidence that
the produced quanta are high energetic.

\begin{figure}[H]
\centerline{\parbox{0.7\textwidth}{\epsfig{file=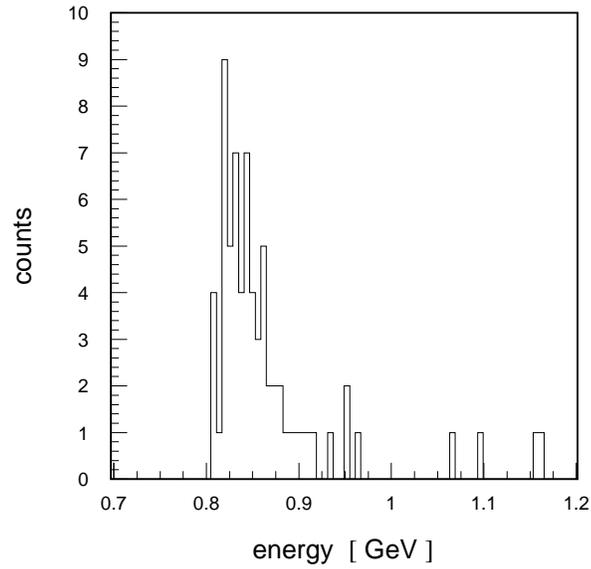,width=0.75\textwidth,angle=0.}}}
      {\caption{ Experimental distribution of gamma energy
                for the $dp~\to~dp\gamma$ reaction. The spectrum
                is not corrected for the detection acceptance.
               }}
\end{figure}

\chapter{Summary and perspectives}
\markboth{\bf Summary and perspectives}
         {\bf Summary and perspectives}

The experiment which was described in this thesis has been performed
at the cooler synchrotron COSY in the Research Center J{\"u}lich
by means of the COSY--11 detection system.\\
The results of the identification of bremsstrahlung radiation
in data taken with a proton target and a deuteron beam
have been presented and discussed. For the first time ---
using COSY-11 facility ---
events associated with the $dp~\to~dp\gamma$ reaction
have been observed. For the quantitative determination of the total
cross section of the $dp~\to~dp\gamma$ reaction the luminosity and
detection acceptance remains to be established. There are also plans
to analyse the data in view of bremsstrahlung radiation
in a quasi--free $np~\to~np\gamma$ reaction.\\
The second point of this work was the time calibration of the neutron detector.
As shown in the chapter 4, the general time offset of neutron detector with respect
to the S1 detector was found to be 13.6 ns. The resolution of the neutron momentum
determination by means of the neutral particle counter --- the crucial factor in
neutron momentum determination --- was found to be 0.14~GeV/c at a neutron momentum of 1.6~GeV/c,
the dependence of the fractional momentum resolution as a function of a neutron momentum
is presented in Appendix B.
\par
The installation of the neutron detector enables not only to study the isospin dependence
of the meson production~\cite{moskal-hadron} and bremsstrahlung radiation, but also
to investigate the production of the resonance $\Theta^{+}$ in the elementary proton--proton
interaction. A signature of the $\Theta^{+}$ production may be the presence of a 1.54~GeV/c$^2$
peak in the $n~K^{+}$ invariant mass distribution for the $pp~\to~nK^{+}\Sigma^{+}$ reaction.
The data analysis aiming for the determination of the n~K$^{+}$ invariant mass distribution has just started.
In order to determine the acceptance of the COSY--11
detection system for the $pp~\to~nK^{+}\Sigma^{+}$
reaction we have simulated the response of the detectors
for $ 6 \cdot 10^{6}$ events generated in the target.
The solid histogram in figure 6.1(left) illustrates the distribution
of the invariant mass $ nK^{+}$ for the generated events
while the dashed line depicts the spectrum which was reconstructed
from the signals simulated in the detectors.

\begin{figure}[H]
\parbox{0.52\textwidth}{\hspace{-0.0cm} \epsfig{file=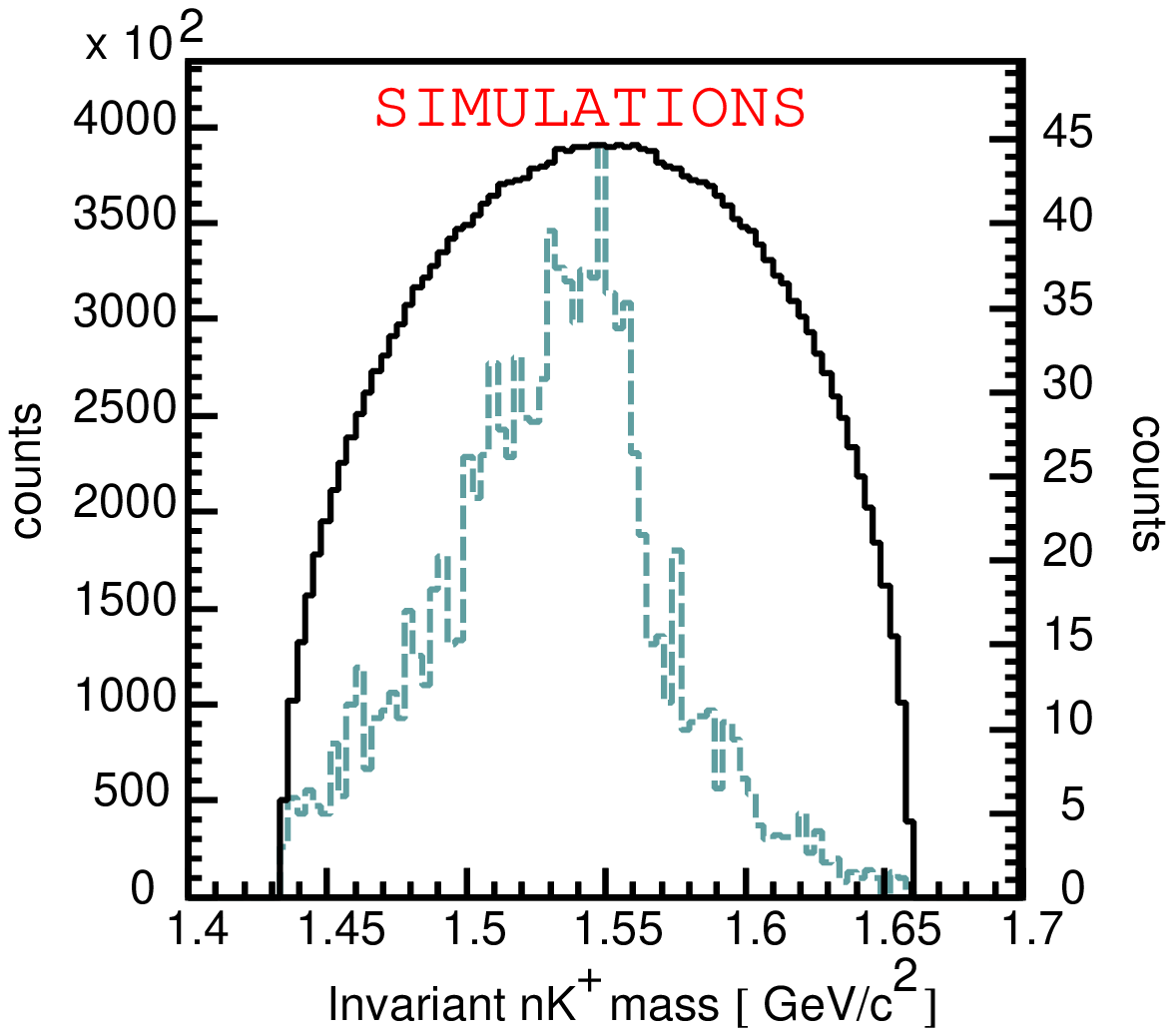,width=0.51\textwidth,angle=0.}} \hfill
\parbox{0.46\textwidth}{ \hspace{0.0cm} \epsfig{file=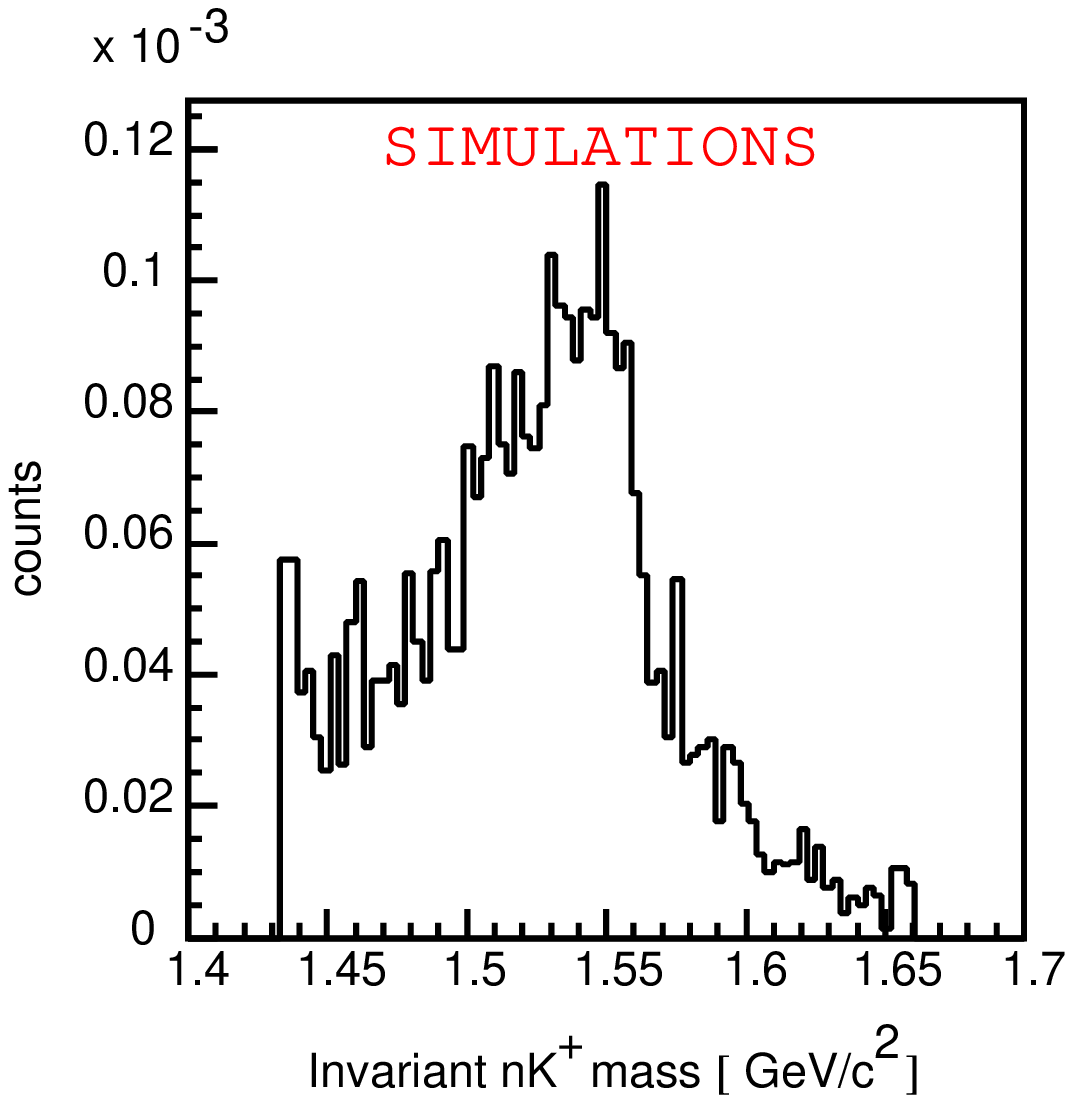,width=0.46\textwidth,angle=0.}}
       {\caption{ {\bf (left)} Phase space distribution of the invariant mass
                  of the $nK^{+}$ system of the $pp~\to~nK^{+}\Sigma^{+}$
                  reaction (solid line) and its convolution with the
                  COSY--11 acceptance (dashed line).
                  {\bf (right)} The acceptance of the COSY--11 detection setup
                  for the $pp~\to~nK^{+}\Sigma^{+}$ reaction~\cite{przerwa-penta}.}}
\end{figure}
The ratio of the obtained invariant mass distributions
results in the differential acceptance of COSY--11 facility for detecting
the $pp~\to~nK^{+}\Sigma^{+}$ reaction as shown in figure 6.1(right).

The total acceptance of COSY--11 detection system for the
$pp~\to~nK^{+}\Sigma^{+}$ reaction measured at the beam momentum of
3.257~GeV/c is equal to $10^{-4}$~\cite{przerwa}.

Additionally, the simulation of the invariant mass distribution
of the process $pp~\to~\Sigma^{+}\Theta^{+}~\to~nK^{+}\Sigma^{+}$
has been performed, taking into account the width of $\Theta^{+}$
equal to 5~MeV~\cite{poliakov}. The results of the Monte--Carlo calculation are
shown in figure 6.2(left).\\

The expected signal from the $pp~\to~\Sigma^{+}\Theta^{+}~\to~n~K^{+}\Sigma{+}$ reaction together
with the background originating from the direct $pp~\to nK^{+}\Sigma^{+}$
reaction is presented in figure 6.2(right).\\
Here it is assumed arbitrarily that the total cross section for the
$pp~\to~\Sigma^{+}\Theta^{+}$ reaction is ten times smaller than the one
for $pp~\to~nK^{+}\Sigma^{+}$. If the performed appraisals
are realistic, one should observed a clear signal in the experiment
originating in the $\Theta^{+}$ production as it is noticeable in the
right panel of figure 6.2.
\begin{figure}[H]
\parbox{0.45\textwidth}{\epsfig{file=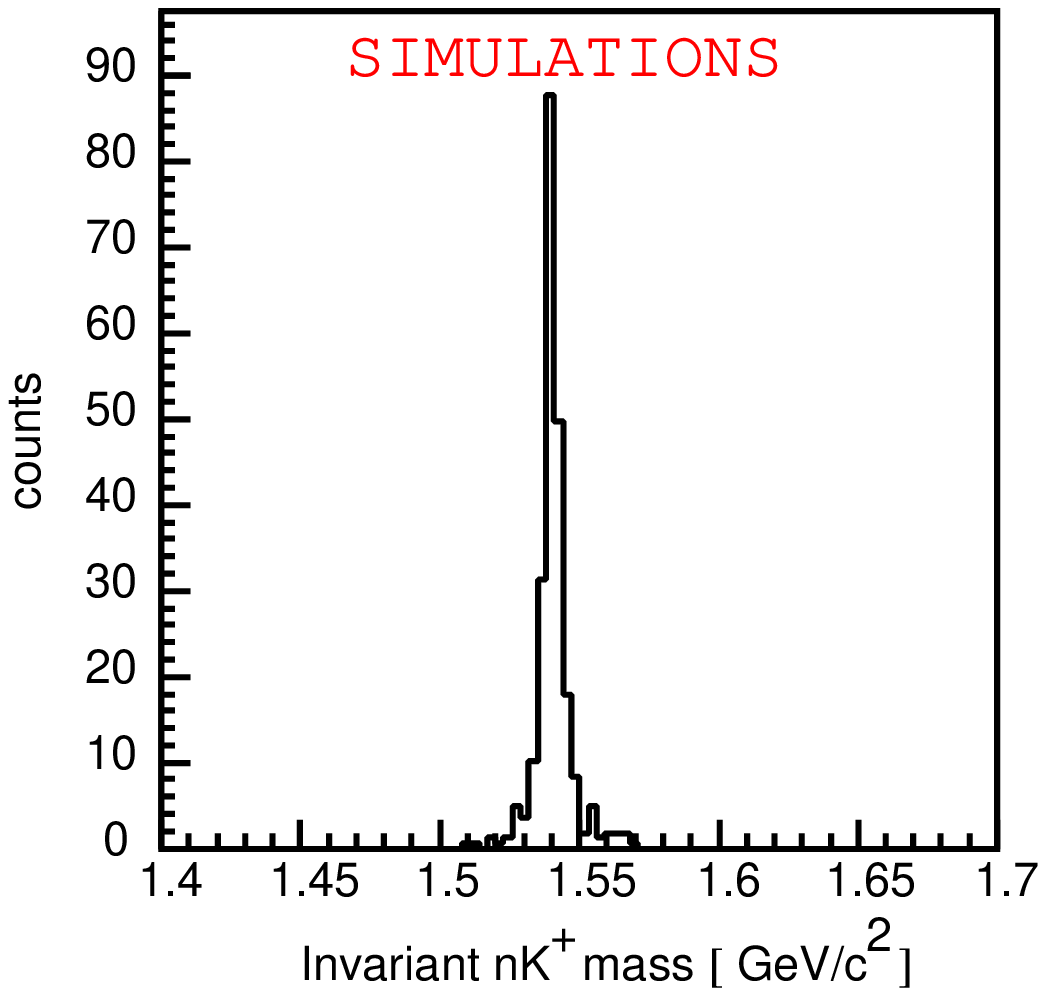,width=0.44\textwidth,angle=0.}} \hfill
\parbox{0.45\textwidth}{\epsfig{file=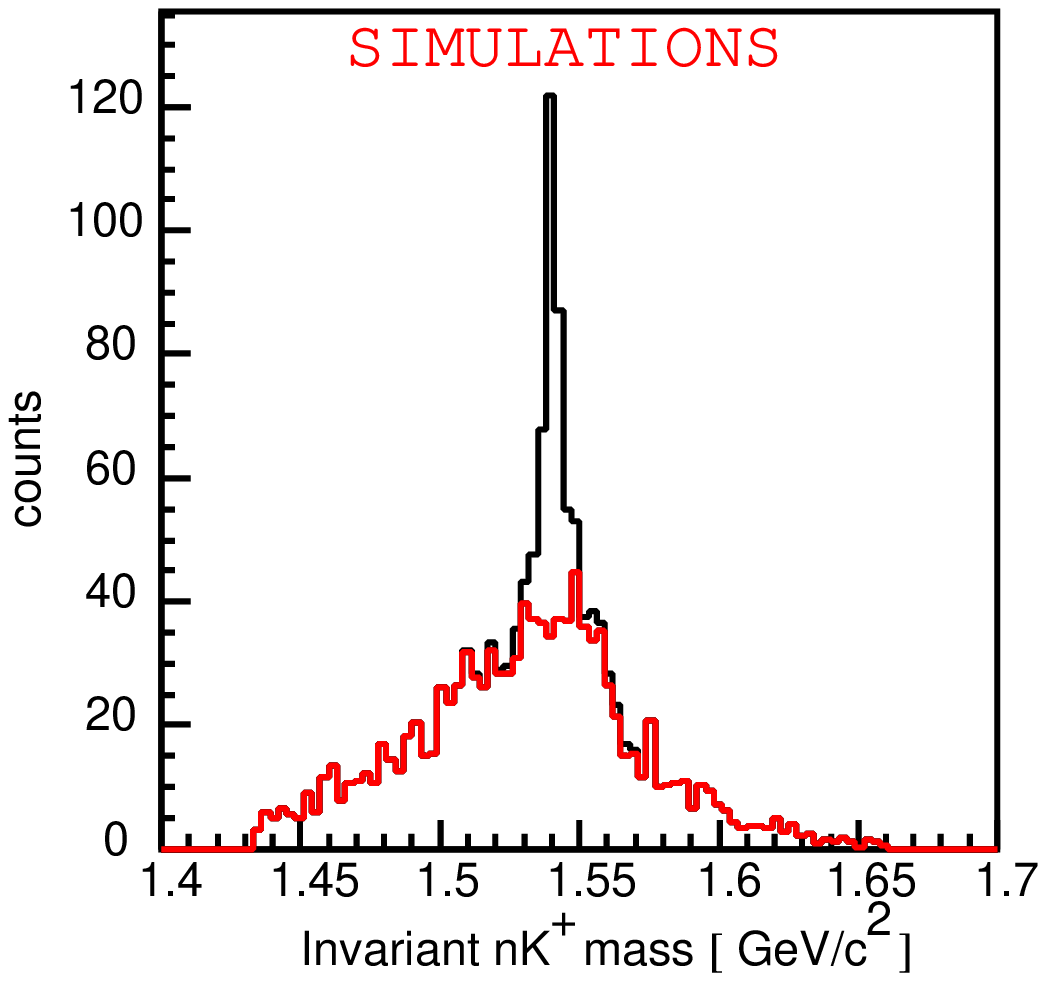,width=0.44\textwidth,angle=0.}}
{\caption{ {\bf (left)} Invariant mass distribution of the $nK^{+}$ system
                  for the $pp~\to~\Sigma^{+}\Theta^{+}~\to~nK^{+}\Sigma^{+}$
                  reaction.
                  {\bf (right)} The expected full invariant mass distribution
                   of the $nK^{+}$ system~\cite{przerwa-penta}.}}
\end{figure}

\chapter*{A. Kinematics of  the $dp~\to~ppn_{sp}$ reaction}
\markboth{\bf Appendix }{\bf Appendix }
\addcontentsline{toc}{chapter}{\protect\numberline{}{A Kinematics of the $dp~\to~ppn_{sp}$ reaction}}

The quasi--free reaction $dp~\to~ppn_{sp}$ has been used to determine
(i) the relative timing between modules, (ii) the general time offset of the neutron detector,
and (iii) to establish the momentum resolution of this detector. Due to the Fermi motion of
the nucleons bound in the deuteron the simulation of the quasi--free reaction
proceeds in following steps :\\
The components of the Fermi momentum of a nucleon in Cartesian coordinate system are equal to:

 $$p_{fx} = p_F \cdot \sin\theta\cos\phi$$
 $$p_{fy} = p_F \cdot \sin\theta\sin\phi$$
 $$p_{fz} = p_F \cdot \cos\theta$$
As a first step the value of p$_F$ (absolute Fermi momentum) is generated according to the momentum distribution
of nucleon inside the deuteron derived from the PARIS potential model~\cite{czyzyk-mgr,lacombe}.
Next the azimuthal angle $\phi$ and cosine of polar angle $\Theta$ ($\cos\Theta$)
which define the momentum direction, are  generated
assuming an uniform distribution. Further the nucleon Fermi momentum inside the deuteron
is related to the nucleon  momentum in laboratory frame by Lorentz transformation~\cite{kajantie}:
$$\vec {p_f}^{lab}=\vec p_f+ \vec \beta_{d} \gamma(\gamma/(\gamma+1) \vec \beta_{d}\cdot \vec p_f + E_f)$$
 $$E_f^{lab}=\gamma (E_f + \vec \beta_{d} \cdot \vec p_f)$$

where $\beta_{d}$ is the velocity of the deuteron in the laboratory frame and $\gamma$ is equal to:
$$\gamma = 1/\sqrt{1- \beta_{d}^2}$$

After the momenta of the nucleons inside the deuteron are converted into the LAB system,
the proton from the deuteron scatters elastically on a proton target, whereas the neutron does
not take part in the reaction,
but with momentum possessed at the time of the reaction remains untouched.
In order to simulate the proton--proton elastic scattering
now we calculate the proton momentum in the proton--proton system using the transformation equation:
$$ |\vec p_{cm}| = \sqrt{(M^2-4m_{p}^2)/4}$$
where M is the total mass of the colliding protons:

$$  M = \sqrt{ (\sqrt {\vec {p_f}^{{lab}^2} + m_p^2} + m_p)^2 - \vec {p_f}^{{lab}^2}}$$

Once more  $\cos\Theta^{cm}$ and $\phi^{cm}$, which define the momentum direction of protons
after the scattering in the proton--proton system are generated assuming an uniform distribution.
Thus, the components of the proton momentun after scattering are equal to:

$$ p_{x}^{cm'} = |\vec{ p^{cm}}| \sin\theta^{cm}\cos\phi^{cm}$$

$$ p_{y}^{cm'} = |\vec p^{cm}|\sin\theta^{cm}\sin\phi^{cm}$$

$$ p_{z}^{cm'} = |\vec p^{cm}|\cos\theta^{cm}$$

As a last step the proton momentum after scattering in the proton--proton system is related with momentum
in laboratory frame by Lorentz transformation:

$$\vec {p^{lab'}} = \vec {p^{cm'}}+\vec {\beta^{cm}}\gamma^{cm}(\gamma^{cm}/(\gamma^{cm} + 1)\vec{\beta^{cm}}\cdot \vec{p^{cm'}} + E^
{cm'}),$$

where\\

 $$\vec { \beta^{cm}} = {{\vec{ p^{lab}_f} } \over { \sqrt{{\vec{ p^{lab}_f} }^2 + m^2_p} + m_p  } },$$

and\\

$$ \gamma^{cm} = {{\sqrt{{\vec{ p^{lab}_f} }^2 + m^2_p} + m_p} \over M}.$$

\chapter*{B. Resolution of the measurement of the neutron momentum}
\markboth{\bf Appendix }{\bf Appendix }

\addcontentsline{toc}{chapter}{\protect\numberline{}{B Resolution of the measurement of the neutron momemtum}}
The momentum resolution of the neutron detector is a crucial factor
in the neutron momentum determination and as shown in figure 6.3 it strongly depends
on the momentum of the neutron.
\begin{figure}[H]
\centerline{\parbox{0.7\textwidth}{\epsfig{file=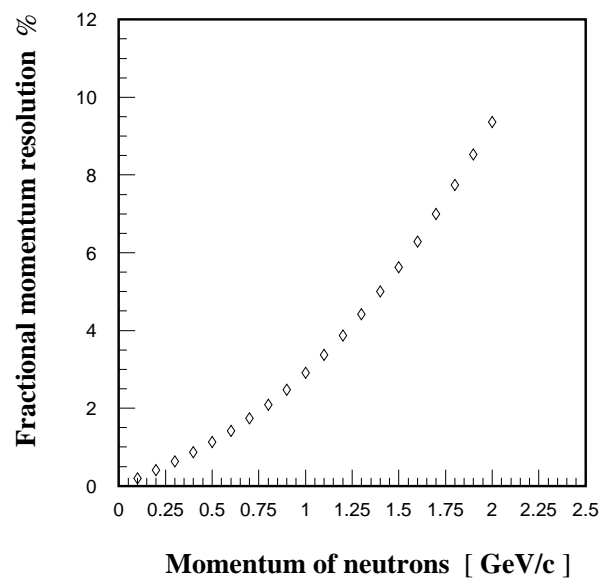,width=0.75\textwidth,angle=0.}}}
      {\caption{  Fractional momentum resolution of the neutron detector
                  as a function of neutron momentum.
               }}
\end{figure}
The momentum of the neutron is calculated from the time--of--flight between the
target and the neutron detector and can be expressed as:

$$ p = m \cdot {l \over {t^N}} \cdot  {1 \over {\sqrt{1 - ({l \over {t^N}})^2}}},$$

where $m$ denotes mass of the particle, $l$ stands for the distance between the target and the
neutron detector and $t^N$ is the time--of--flight of the particle.
A fractional momentum resolution is given by the equation:

$$  {{\sigma p} \over p} = {{{{dp} \over {dt}} \cdot \sigma_t} \over p},$$

where $\sigma_t$ accounts for both the time resolution of the neutron and S1 detectors and can be
written as:

 $$ \sigma_t = \sqrt {\sigma_n^2 + \sigma_{S1}^2}$$

The result presented in figure 6.3 was obtained assuming that $\sigma_n$ =~0.4~ns, $\sigma_{S1}$~=~0.25~ns,
and $l$~=~7~m.

\chapter*{C. Dalitz plot distribution for the $dp\gamma$ system at $\sqrt s$~=~3.37~GeV}
\markboth{\bf Appendix }{\bf Appendix }
\addcontentsline{toc}{chapter}{\protect\numberline{}{C Dalitz plot distribution for the $dp\gamma$ system at $\sqrt s $~=~3.37~GeV}}

As a first step for establishing of the double differential
acceptance of the COSY--11 facility for the measurement of
the $dp~\to~dp\gamma$ reaction a Dalitz plot distribution for the $dp~\to~dp\gamma$ reaction
was calculated. Figure 6.4 presents the result of calculations assuming
that the phase space volume is homogeneously populated.
Modifications of this distribution due to the COSY--11 detection setup acceptance remain to
be determined. \\

\begin{figure}[H]
\centerline{\parbox{0.6\textwidth}{\epsfig{file=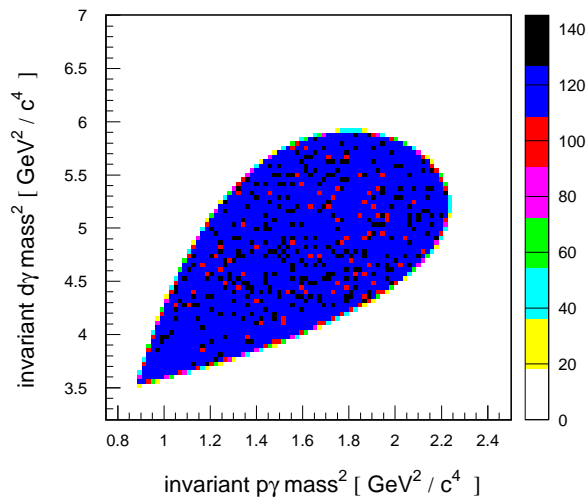,width=0.65\textwidth,angle=0.}}}
      {\caption{
                    Dalitz plot distribution as generated for  the $dp~\to~dp\gamma$
                    reaction.
      }}
\end{figure}
Squared of the invariant masses of the $p\gamma$ and $d\gamma$ system can be calculated
according to the below equations:

$$S_{p\gamma} = {(E_{p} + E_{\gamma})}^2 - {( \vec {p_{p}} + \vec {p_{\gamma}} )}^2$$

$$S_{d\gamma} = {(E_{d} + E_{\gamma})}^2 - {( \vec {p_{d}} + \vec {p_{\gamma}} )}^2$$

It is interesting to note that minimum values for the S$_{p\gamma}$ and S$_{d\gamma}$ are equal to:

$$S_{p\gamma}^{min} = M_{p}^2$$

$$ S_{d\gamma}^{min} = M_{d}^2$$

\newpage
\clearpage
\pagestyle{plain}
\pagestyle{myheadings}
\markboth{\bf References }{\bf References }
\addcontentsline{toc}{chapter}{\protect\numberline{}{References}}

\end{document}